\documentclass[
twocolumn,
aps,
prb,
reprint,
superscriptaddress,
amsmath,amssymb
]{revtex4-1}
\usepackage[pdftex]{graphicx} 
\usepackage[hidelinks]{hyperref} 
\hypersetup{
 setpagesize=false,
 bookmarksnumbered=true,%
 bookmarksopen=true,%
 colorlinks=true,%
 linkcolor=blue,
 citecolor=blue,
}

\usepackage{braket}
\usepackage{times,multirow,amsfonts,bm,xspace,pifont}
\usepackage{soul}
\usepackage{ulem}
\usepackage{mathrsfs}
\begin{document}
\newcommand{\rr}{{\bm r}}
\newcommand{\q}{{\bm q}}
\renewcommand{\k}{{\bm k}}
\newcommand*\wien    {\textsc{wien}2k\xspace}
\newcommand*\textred[1]{\textcolor{red}{#1}}
\newcommand*\textblue[1]{\textcolor{blue}{#1}}
\newcommand*\YY[1]{\textcolor{blue}{#1}}
\newcommand*\JI[1]{\textcolor{red}{#1}}
\newcommand*\AD[1]{\textcolor{red}{#1}}
\newcommand*\ADS[1]{\textcolor{blue}{\sout{#1}}}
\newcommand*\ADC[1]{\textcolor{green}{#1}}
\newcommand{\KT}[1]{{\color{red}{#1}}}
\newcommand\nazo[1]{\textcolor{red}{#1}}

\title{Thermodynamic electric quadrupole moments of nematic phases from first-principles calculation} 
\author{Taisei Kitamura}
\email[]{kitamura.taisei.67m@st.kyoto-u.ac.jp}
\affiliation{Department of Physics, Graduate School of Science, Kyoto University, Kyoto 606-8502, Japan}

\author{Jun Ishizuka}
\email[]{ishizuka.jun.8c@kyoto-u.ac.jp}
\affiliation{Department of Physics, Graduate School of Science, Kyoto University, Kyoto 606-8502, Japan}

\author{Akito Daido}
\email[]{daido@scphs.kyoto-u.ac.jp}
\affiliation{Department of Physics, Graduate School of Science, Kyoto University, Kyoto 606-8502, Japan}

\author{Youichi Yanase}
\email[]{yanase@scphys.kyoto-u.ac.jp}
\affiliation{Department of Physics, Graduate School of Science, Kyoto University, Kyoto 606-8502, Japan}
\affiliation{%
  Institute for Molecular Science, Okazaki 444-8585, Japan
}
\date{\today}

\begin{abstract}
The electronic nematic phase emerging with spontaneous rotation
symmetry breaking is a central issue
of modern condensed matter physics.
In particular,
various nematic phases 
in iron-based superconductors and high-$T_{\rm c}$ cuprate superconductors
are extensively studied recently.
Electric quadrupole moments (EQMs) are
one of the order parameters characterizing these nematic
phases in a unified way, and
elucidating EQMs is a key to
understanding these nematic phases.
However, the quantum-mechanical formulation of the EQMs in crystals is a nontrivial issue because the position operators are non-periodic and unbound. Recently, the EQMs have been formulated by local thermodynamics, and such {\it thermodynamic EQMs} may be used to characterize the fourfold rotation symmetry breaking in materials.
In this paper,
we calculate the thermodynamic EQMs in iron-based superconductors LaFeAsO and FeSe as well as a cuprate superconductor La$_2$CuO$_4$ by a first-principles calculation. 
We show that owing to the orbital degeneracy the EQMs in iron-based superconductors are mainly determined by the geometric properties of wave functions. This result is in sharp contrast to the cuprate superconductor, in which the EQMs are dominated by distortion of the Fermi surface. 
\end{abstract}

\maketitle

\section{Introduction}
In recent years, the nematic phases 
which spontaneously break fourfold rotation ($C_4$) symmetry
are attracting a lot of interest in condensed matter physics.
For example, iron-based superconductors such as LaFeAsO\cite{Cruz2008,LaFeAsO_Nomura}, FeSe\cite{FeSe_Hsu,FeSe_margadonna,FeSe_Bohmer}, and BaFe$_2$As$_2$\cite{BaFe2As2,BaFe2As2_Huang} undergo nematic order with an electronic origin, although it is 
accompanied by the tetragonal-orthorhombic structural phase transition. 
The relations of nematic order, superconductivity, and magnetism have been of central interest in the research of iron-based superconductors in the past decade
\cite{Ishida2009,Stewart2011,Dai2015,Shibauchi2020,mazin2008unconventional,kuroki2008unconventional,ikeda2008pseudogap,wang2008superconductivity,yanagi2010orbital,thomale2011exotic,Kontani_nematic,Onari2012,Fernandes2010,Fernandes2012,Yamada2014Fe,Mukherjee2015}.
The nematic order in cuprate superconductors is also a topic of interest, motivated by recent experimental indications\cite{Daou2010,sato2017thermodynamic}. Although vast studies have been devoted, comprehensive clarification of the nematic order and its relation to the pseudogap phase and superconductivity remains an ongoing issue. 
For an origin of the nematic order in cuprates, the charge density wave (CDW) order and bond order~\cite{Yamase2000,Halboth2000,Honerkamp2001,Kee2004,berg2009striped,Metzner2012,bulut2013spatially,sachdev2013bond,wang2014charge,yamakawa2015spin,kawaguchi2017competing,tsuchiizu2018multistage} as well as other exotic order such as loop current and pair density wave~\cite{affleck1988large,schlz1989fermi,zhang1990superconducting,varma1997non,lee2014amperean,agterberg2020the} have been proposed. 
Because of a possible interplay with the high-$T_{\rm c}$ superconductivity in the two categories, namely, iron-based and cuprate superconductors, the nematic phase in strongly correlated electron systems
is a central issue of modern condensed matter physics.
In particular, the similarities and differences between iron-based superconductors and cuprate superconductors are issues to be solved. 

Although various nematic order parameters have been proposed, a ubiquitous feature is the $C_4$-symmetry breaking. 
The electric quadrupole moments (EQMs) are one of the fundamental quantities which naturally characterize the $C_4$-symmetry breaking~\cite{Jackson}. 
Therefore studying EQMs is a key to understanding the nematic phases, and may also give an insight into the relations to superconductivity.

EQMs are originally introduced in classical electromagnetism~\cite{Jackson}.
However, contrary to their apparently simple form, a naive extension to periodic crystals is problematic due to the difficulties in the treatment of the position operator.
This is also the case for magnetic multipole moments.
In contrast to the modern theory of electric polarization where the approach based on Wannier functions is successful\cite{vanderbilt_2018,dp_king,dp_vander}, EQMs of the Wannier functions are gauge dependent unless appropriate symmetries are preserved~\cite{Marzari1997,Marzari2012}. On the other hand, EQMs have been well formulated as topological invariants of higher order topological crystalline insulators\cite{BBH_science,BBH_prb,Fe_topeqm,Ezawa_topeqm,Imhof_topoeqm,Serra_topoeqm,Zhida_topeqm,Franca_topeqm,Hirayma_topoeqm,Peterson_topoeqm,Watanabe_topeqm,Schindler_topeqm,Becalcazar_topeqm}.
However, obtained results are valid
only in the presence of 
crystalline symmetries such as the $C_4$ symmetry. 
Therefore, these approaches are not useful for evaluating EQMs that emerge with spontaneous $C_4$-symmetry breaking.

Although the above quantum mechanical approach is still an ongoing issue, 
recent thermodynamic approaches to the  electric/magnetic multipole moments\cite{shitade_mqm,shitade_mqm2,gao2018microscopic,gao2018orbital,TEQMs} successfully obtained
gauge-invariant and unit-cell independent formulas.
In particular, the EQMs obtained by the thermodynamic approach, which we call thermodynamic EQMs, are well defined even without crystalline symmetries such as
the $C_4$ symmetry, in contrast to the EQMs formulated as topological invariants.
Therefore, using the thermodynamic EQMs, now we can study the nematic phases of iron-based superconductors and cuprate superconductors in a unified manner.

In this paper,
after
showing the failures of the formulations 
by electromagnetism and 
Wannier function methods,
we calculate the thermodynamic EQMs 
of two iron-based superconductors,
LaFeAsO, FeSe, and a cuprate superconductor 
La$_2$CuO$_4$ using 
first-principles calculations.
For the nematic order parameters, the orbital order and momentum-dependent orbital polarization are assumed for LaFeAsO and FeSe, respectively, in accordance with theoretical proposals~\cite{Kontani_nematic,Onari2012,yanagi2010orbital,ohnari_sign}. For La$_2$CuO$_4$, we examine the $d_{x^2-y^2}$-wave bond order as well as the orbital order of O2$p_{x}$ and O2$p_{y}$ orbitals for a comparison. 
The results reveal a difference in the EQMs between the iron-based superconductors and cuprate superconductors.
The Fermi-sea term in the thermodynamic EQM is dominant in iron-based superconductors, owing to 
the unique band structures and associated geometric properties. 
In contrast, the thermodynamic EQM in the cuprate superconductor is dominated by the Fermi-surface term arising from the distortion of the band structure. 

\section{Electromagnetic EQMs in crystals}

We begin with the definition of EQMs in classical electromagnetism. 
With the multipole expansion, 
the scalar potential is described as\cite{Jackson},
\begin{eqnarray}
\phi(\bm r) =\dfrac{1}{4\pi \varepsilon}
\int d\bm r^\prime \sum_{l = 0}^\infty
\rho(\bm r^\prime)
\dfrac{r^{\prime l}}
{r^{l+1}}P_l(\hat{\bm r}\cdot \hat{\bm r}^\prime).
\end{eqnarray}
Here,
$\varepsilon$ is the dielectric constant,
$P_l(x)$ are the Legendre polynomials,
$\hat{\bm r}$ is the unit vector of $\bm r$, i.e. 
$\hat{\bm r} = \bm r /|\bm r|$, 
and $\rho(\bm r)$ is the charge density.
For simplicity, we adopt the units with electric charge $e = 1$ and lattice volume $V_{\text{cell}} = 1$.
We focus on the component of $l = 2$, 
which is represented as 
\begin{eqnarray}
    &&\phi^{(2)}(\bm r)
    =\dfrac{1}{4\pi \epsilon}
    \sum_{ij}
    \dfrac{1}{2r^5}
    (4r_ir_j
    -r^2\delta_{ij}) \mathcal{Q}_{ij},
\end{eqnarray}
with the electromagnetic EQMs 
\begin{eqnarray}
    &&\mathcal{Q}_{ij}
    = \int
    d{\bm r}r_ir_j\rho(\bm r).
    \label{emEQMs}
\end{eqnarray}
The electromagnetic EQMs are determined by the quadrupole distribution of the charge density. 
The other multipole moments are defined as well in a similar manner.

Here, we try to evaluate the electromagnetic EQMs of electrons on the periodic crystal lattice. 
Difficulties due to the non-periodic and unbound position operators will become manifest.
We consider the spinless case for simplicity. 
An idea to avoid the unboundedness of the position operator, which can be infinite in the thermodynamic limit, is to consider the EQMs in a unit cell. The EQMs are redefined as
\begin{eqnarray}
    \mathcal{Q}_{ij}^{\rm cell}
    &=&\int_{\rm cell}dr^d
    r_ir_j
    \braket{\psi^\dagger(\bm r)
    \psi(\bm r)}\notag\\
    &=& \dfrac{1}{V}\sum_{nm}\sum_{\bm k \bm k^\prime}
     \int_{\rm cell}d\bm r
    r_ir_j
    \braket{c_{n}^\dagger(\bm k)
    c_{m}(\bm k^\prime)}\notag\\
    &&\times
    e^{-i(\bm k-\bm k^\prime)\cdot\bm{r}}
    u_{n}^{*}(\bm{k},\bm r)
    u_{m}(\bm{k}^\prime,\bm r),
\end{eqnarray}
with 
$\psi^\dagger(\bm r)$
being the creation 
operator of electrons at $\bm r$. 
In the second line,
we expanded $\psi(\bm r)$ by the periodic part of the Bloch wave function $u_{n}(\bm{k},\bm r)$:
$\psi(\bm r) = \dfrac{1}{\sqrt{V}}\sum_{n}\sum_{\bm k}c_{n}(\bm k)e^{i\bm{k}\cdot\bm{r}}u_{n}(\bm{k},\bm r)$. 
Here, $m,n$ are band indices, $\bm k$ is the wave vector, $V$ is the volume of the system, and $d$ is the dimension of the system.
The momentum sum is carried out in the first Brillouin zone. 
In this notation, $u_n(\bm k,\bm r)$ are normalized as $\int dr^d u_m^*(\bm k,\bm r) u_n(\bm k,\bm r) = \delta_{nm}$.
For free electron systems with band dispersion $\epsilon_n(\bm k)$, the relation 
\begin{eqnarray}
    \braket{c_{n}^\dagger(\bm k)
    c_{m}(\bm k^\prime)}
    =f(\epsilon_n(\bm k))
    \delta_{nm}\delta_{\bm k,\bm k^\prime},
\end{eqnarray}
leads to 
\begin{eqnarray}
    \mathcal{Q}_{ij}^{\rm cell} = \sum_{n}\int_{\rm BZ}\dfrac{dk^d}{(2\pi)^d}
    \int_{\rm cell}dr^d r_ir_j
    f(\epsilon_n(\bm k))
    u_{n}^*(\bm{k},\bm r)
    u_{n}(\bm{k},\bm r).\notag\\
\label{EQM_unitcell}
\end{eqnarray}
Here, $f(\epsilon)=(e^{\epsilon/T}+1)^{-1}$ represents the Fermi distribution function for the temperature $T$.

\begin{figure}[htbp]
    \includegraphics[width=0.9\linewidth]{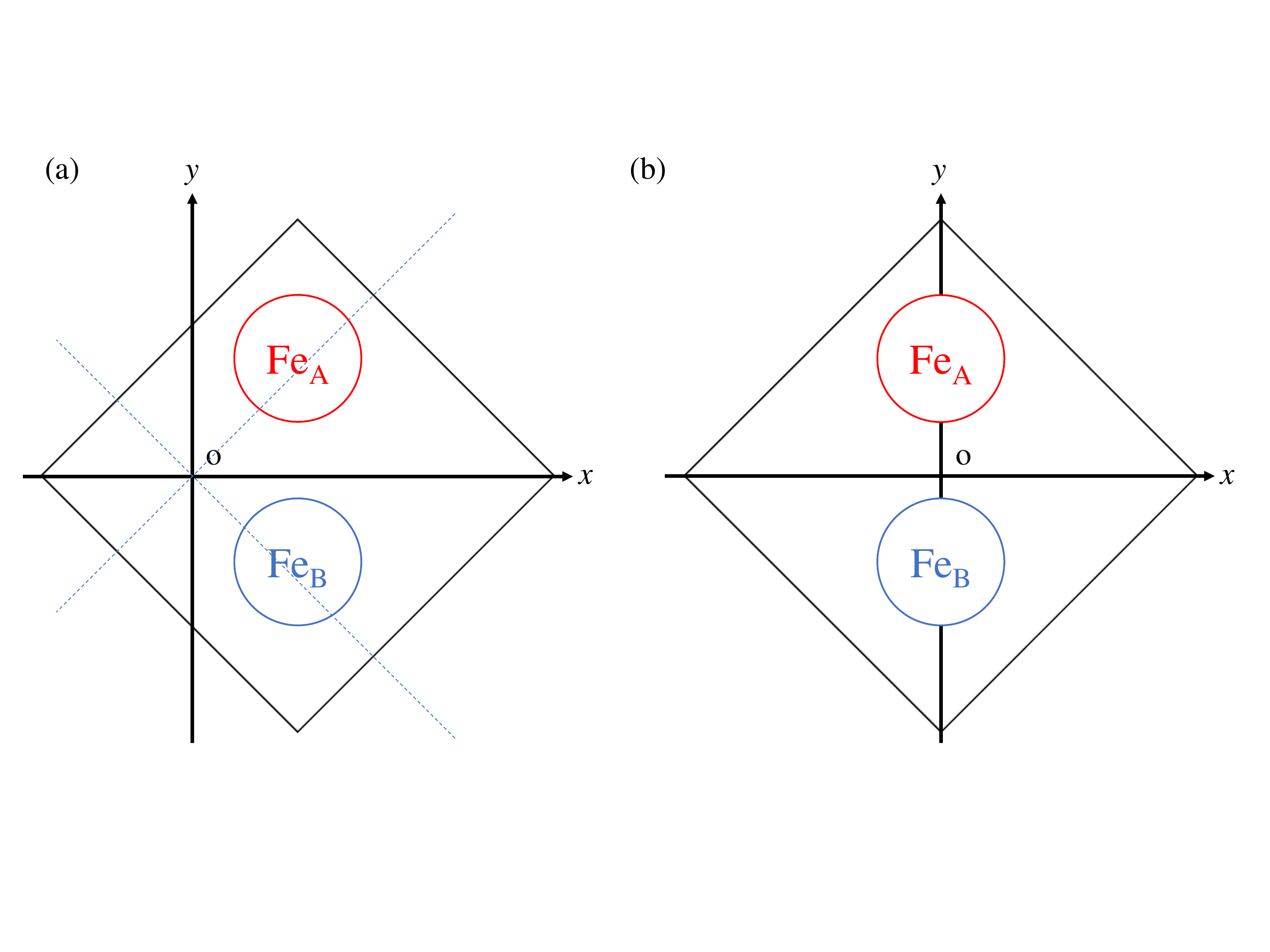}
    \centering
    \caption{Unit cell of LaFeAsO and choices of real space coordinates.
    For the case (a), the electromagnetic EQM,\ $\mathcal{Q}_{x^2-y^2} = \mathcal{Q}_{x^2}-\mathcal{Q}_{y^2}$, vanishes,
    while $\mathcal{Q}_{x^2-y^2}$ is finite for the case (b).
    Thus, the EQMs of the unit cell are not well defined.}
    \label{fig:position_sheme}
\end{figure}

The EQMs of the unit cell, Eq.~\eqref{EQM_unitcell}, depends on the origin of real space coordinates. For concreteness, we consider the two-sublattice systems as for LaFeAsO we study in Sec. \ref{sec:teqms}.  
In a tight-binding model,
the orbitals of electrons
are localized on atoms and we have a simple relation
\begin{align}
    &u_{n}^*(\bm{k},\bm r)u_{n}(\bm{k},\bm r)
    = \braket{ u_n(\bm k)\vert \bm r}\braket{\bm r\vert u_n(\bm k)}\notag\\
    &=\sum_{\bm{R}}\sum_{\text{sub}=A,B}|\braket{\bm{r}_{\text{sub}}|u_n(\bm k)}|^2\delta(\bm r- \bm{R}-\bm r_{\rm sub}), 
\end{align}
where $\bm{r}_A,\bm{r}_B$ are the sublattice positions within a unit cell and $\bm{R}$ are the lattice points.
Only the home unit cell $\bm{R}=0$ contributes to the integral~\eqref{EQM_unitcell}.

Examples of the coordinates are shown in Fig.~\ref{fig:position_sheme}.
A EQM characterizing the nematic order is the inbalance
between the $x$ and $y$ directions: $\mathcal{Q}_{x^2-y^2}^{\rm cell}= \mathcal{Q}_{x^2}^{\rm cell}-\mathcal{Q}_{y^2}^{\rm cell}$.
This vanishes for the coordinates in Fig.~\ref{fig:position_sheme}(a), since Fe$_1$ and Fe$_2$ are on the lines of $x^2-y^2 = 0$, while $\mathcal{Q}_{x^2-y^2}^{\rm cell}$ is finite for the case of Fig.~\ref{fig:position_sheme}(b). 
Indeed, we obtain $\mathcal{Q}_{x^2-y^2}^{\rm cell} \simeq -0.75$ at $T=0.1$ for LaFeAsO in the absence of the nematic order, when we choose coordinates in  Fig.~\ref{fig:position_sheme}(b). 
Because of these undesirable properties, (1) $\mathcal{Q}^{\rm cell}_{ij}$ depend on the coordinates, and (2) $\mathcal{Q}^{\rm cell}_{ij}$ can be finite even in the absence of the nematic order, the electromagnetic EQMs of the unit cell are not suitable for quantifying the nematic order. 

We also discuss the EQMs of Wannier functions defined by the moment of position operators:
\begin{align}
    &\mathscr{Q}^{(n)}_{ij} = 
    \bra{W_{\bm 0 n}}\hat{r}_i
    \hat{r}_j\ket{W_{\bm 0 n}}\notag\\
    &=-\sum_{m(\neq n)}
    \int_{\rm BZ}\dfrac{dk^d}{(2\pi)^d}
    \braket{u_n(\bm k)\vert
    \partial_{k_i}u_m(\bm k)}
    \braket{u_m(\bm k)\vert
    \partial_{k_j}u_n(\bm k)}
    \notag\\
    &-\int_{\rm BZ}\dfrac{dk^d}{(2\pi)^d}
    \braket{u_n(\bm k)\vert
    \partial_{k_i}u_n(\bm k)}
    \braket{u_n(\bm k)\vert
    \partial_{k_j}u_n(\bm k)},
    \label{Wannier_eqm}
\end{align}
with the Wannier function of the $n$-th band, 
\begin{eqnarray}
    \ket{W_{\bm Rn}} = \int \dfrac{dk^d}{(2\pi)^d}e^{-i\bm k \cdot (\bm R-\hat{\bm r})}\ket{u_n(\bm k)}.
\end{eqnarray}
Clearly,
the second term of Eq.~\eqref{Wannier_eqm}
is gauge dependent.
Therefore, the EQMs of Wannier functions are unsuitable for
evaluating the EQMs of nematic phases.

Instead of the above quantum-mechanical approaches, we adopt the thermodynamic approach as we explain in the next section. 
Note that
the first term of Eq.~\eqref{Wannier_eqm}
appears in the thermodynamic EQMs\cite{TEQMs}.
It is a part of the geometric term and given by the quantum metric\cite{q_met_Resta,provost1980riemannian}, namely, the real part of the quantum geometric tensor.

\section{Thermodynamic EQMs}
In this section, we introduce the thermodynamic EQMs.
The EQMs are recently formulated by the variation 
of the free energy density\cite{TEQMs}:
\begin{widetext}
\begin{align}
    dF(\bm r) = \rho(\bm r)d\phi(\bm r) + p_i d[\partial_i\phi(\bm r)]
    + Q_{ij} d[\partial_i\partial_j\phi(\bm r)]
    + \mathcal{O}(d[\nabla^3\phi(\bm r)],[d\phi(\bm r)]^2).
\label{freeenergy}
\end{align}
\end{widetext}
While the charge density $\rho(\bm r)$ defined by the differential with respect to the scalar potential $\phi(\bm r)$ is regarded as thermodynamic electric monopole moment, the electric dipole $p_i$ and quadrupole moments $Q_{ij}$ are given by the differential with respect to the spatially nonuniform scalar potential. 
The thermodynamic EQMs are defined as the change of free energy density by the nonuniform electric field, and therefore, it is naturally related to the quadrupole charge distribution.

The expressions for the thermodynamic EQMs $Q_{ij}$ are
obtained as~\cite{TEQMs}
\begin{widetext}
\begin{eqnarray}
    Q_{ij} = \sum_n \int_{\rm BZ}\dfrac{d^dk}{(2\pi)^d}
    \left[
        \dfrac{1}{2}g_n^{ij}(\bm k)f(\epsilon_n(\bm k)) 
        -X^{ij}_n(\bm k)\int^\infty_{\epsilon_n(\bm k)}d\epsilon f(\epsilon)
        -\dfrac{1}{12}m_n^{-1}(\bm k)^{ij}f^\prime(\epsilon_n(\bm k))
    \right].\label{eq:TEQMs}
\end{eqnarray}
\end{widetext}
Here, $g_n^{ij}(\bm k)$ is the quantum metric\cite{q_met_Resta,provost1980riemannian}:
\begin{eqnarray}
    g_n^{ij}(\bm k) = \sum_{m\neq n}A_{nm}^i(\bm k)A_{mn}^j(\bm k) + {\rm c.c.},\label{eq:quantum_metric}
\end{eqnarray}
which is a counter-part of the Berry curvature. 
The quantum metric is the real part of the quantum geometric tensor~\cite{q_met_Resta,provost1980riemannian}, while the imaginary part is the Berry curvature. Equation~\eqref{eq:TEQMs} reveals that the momentum integral of the quantum metric gives a part of the thermodynamic EQMs, while the integral of the Berry curvature is an intrinsic part of the anomalous Hall conductivity~\cite{XiaoRMP2010}.  
In the second term, $X_n^{ij}(\bm k)$ is given by
\begin{eqnarray}
    X_n^{ij}(\bm k) = 
    -\sum_{m\neq n}\dfrac{A_{nm}^i(\bm k)A_{mn}^j(\bm k) + {\rm c.c.}}
    {\epsilon_n(\bm k)-\epsilon_m(\bm k)},\label{eq:X_n}
\end{eqnarray}
while $m^{-1}_n(\bm k)^{ij} = \partial_{k_i}\partial_{k_j}\epsilon_n(\bm k)$
in the third term is the inverse effective mass tensor.
We adopted the following notations
\begin{eqnarray}
    (\hat{H}_0-\mu)\ket{\psi_n(\bm k)} = \epsilon_n(\bm k)\ket{\psi_n(\bm k)},
    \\
    \ket{u_n(\bm k)} = e^{-i\bm k \bm r}\ket{\psi_n(\bm k)},\\
    A_{nm}^i(\bm k) = -i\braket{u_n(\bm k)\vert\partial_{k_i}u_m(\bm k)}.
\end{eqnarray}
$\hat{H}_0$ and $\mu$ are the noninteracting Hamiltonian and the chemical potential, respectively. 
Note that Eq.~\eqref{eq:TEQMs} is valid only when all bands are isolated. General expressions valid in the presence of band touchings are provided in Ref.~\onlinecite{TEQMs}.
Although we adopt single-particle Hamiltonian, the expressions can also be used for the many-body states after the mean-field approximation.

The EQMs formulated based on the thermodynamics are gauge-invariant and independent of unit-cell choices. Thus, difficulties of the electromagnetic EQMs in periodic crystals have been solved. 
Therefore, we evaluate the thermodynamic EQMs in the representative nematic phases of cuprates and iron-based superconductors and discuss their origin.

In Eq.~\eqref{eq:TEQMs}, the first two terms are the contribution from the Fermi sea.
These terms reflect the geometric properties of the Bloch electrons~\eqref{eq:quantum_metric} and \eqref{eq:X_n}, and thus, they are called the Fermi-sea terms in the following.
On the other hand, the third term is a property of Fermi surfaces and vanishes in insulators at zero temperature.
This term is given by the anisotropy of the inverse effective mass tensor on Fermi surfaces and called the Fermi-surface term.
In the following sections, we discuss the difference between iron-based superconductors and cuprate superconductors from the perspective of the thermodynamic EQMs, based on the decomposition into the Fermi-sea and Fermi-surface contributions.
For more details of Eq.~\eqref{eq:TEQMs}, please see Ref.~\onlinecite{TEQMs}.

\section{Thermodyanmic EQMs from first-principles calculation\label{sec:teqms}}
In this section, we calculate and discuss the thermodynamic EQMs
using the first-principles calculation. 
Here, we focus on three representative high-$T_{\rm c}$ superconductors, LaFeAsO, FeSe, and La$_2$CuO$_4$.
The first-principles electronic structure calculations are performed with using the \textsc{wien}2k code \cite{blaha_2}, and the tight-binding models based on the maximally localized Wannier functions \cite{Marzari1997,Souza2001} are constructed by the \textsc{wannier90} code \cite{Mostofi2008}.

The tight-binding Hamiltonian is represented as
\begin{eqnarray}
    H_{\rm LDA} = \sum_{\bm k}\sum_{l,m,i,j,\sigma}t_{lmij}(\bm k)c^\dagger_{li\sigma}(\bm k)c_{mj\sigma}(\bm k), 
    \label{eq:H0}
\end{eqnarray}
where $c^{\dagger}_{li\sigma}(\bm k)$ $(c_{li\sigma}(\bm k) )$ is the creation (annihilation)
operator of the electrons with wave vector $\bm k$, orbital $l$, sublattice $i$, and spin $\sigma$.
The matrix elements $t_{lmij}(\bm k)$ are given by the Fourier transform of the hopping integrals, which are obtained from the \textsc{wien}2k code. Here, we neglect the spin-orbit coupling for simplicity, and evaluate the EQMs per spin. 
To calculate the EQMs in the nematic phases we take into account phenomenological molecular fields for the nematic order parameter. The total Hamiltonian is
\begin{eqnarray}
    H = H_{\rm LDA} + \Delta(T) \Gamma, 
    \label{eq:H}
\end{eqnarray}
where $\Delta(T)$ is the order parameter at the temperature $T$, and $\Gamma$ is the molecular field of the bond order, orbital order, and so on. We specify $\Gamma$ for each compound later. 
To calculate the temperature dependence of the EQMs,
we assume  
\begin{eqnarray}
    \Delta(T) = \left\{
    \begin{array}{cc}
         0 &(T > T_s)\\
          \Delta_0\sqrt{1-T/T_s} &(T\leq T_s),
    \end{array}\right.\label{eq:DeltaT}
\end{eqnarray}
where $T_s$ is the phase transition temperature of the nematic order accompanied by $C_4$-symmetry breaking.
We set $\Delta_0 = T_s = 0.1$, roughly in accordance with the nematic transition temperatures.
Because all materials studied in this paper have quasi-two-dimensional electronic structures, we ignore the hopping integral along the $z$-direction. Thus, two-dimensional multi-orbital and multi-sublattice tight-binding models are analyzed. 
Two EQMs, namely, $Q_{x^2-y^2}$ and $Q_{xy}$ can be finite in two-dimensional systems. In the following, we consider the nematic phases accompanied by a finite $Q_{x^2-y^2}$.
Note that $Q_{xy} = 0$ because of the mirror symmetry.

\subsection{LaFeAsO}

Here we evaluate the EQMs
in LaFeAsO.
We construct 10-orbital tight-binding models for Fe3$d$ electrons. The presence of two iron atoms in a unit cell doubles the  number of orbitals, as $5 \times 2 = 10$. 
It has been suggested that the origin of the nematic order in LaFeAsO is the orbital order of $3d_{xz}$ and $3d_{yz}$ electrons of irons~\cite{Kontani_nematic,Onari2012,yanagi2010orbital}.
Thus, the molecular field is assumed as 
\begin{eqnarray}
\Gamma=\sum_{\bm k}\sum_{\sigma,i=1,2}
    \left[c^\dagger_{d_{{ xz}} i \sigma}(\bm k)c_{d_{{ xz}} i \sigma}(\bm k)  
    -c^\dagger_{d_{{ yz}} i \sigma}(\bm k)c_{d_{{ yz}} i \sigma}(\bm k) \right], 
\notag\\
\end{eqnarray}
where $d_{xz}$ and $d_{yz}$ denote atomic orbitals.

\begin{figure}[htbp]
    \includegraphics[width=1.0\linewidth]{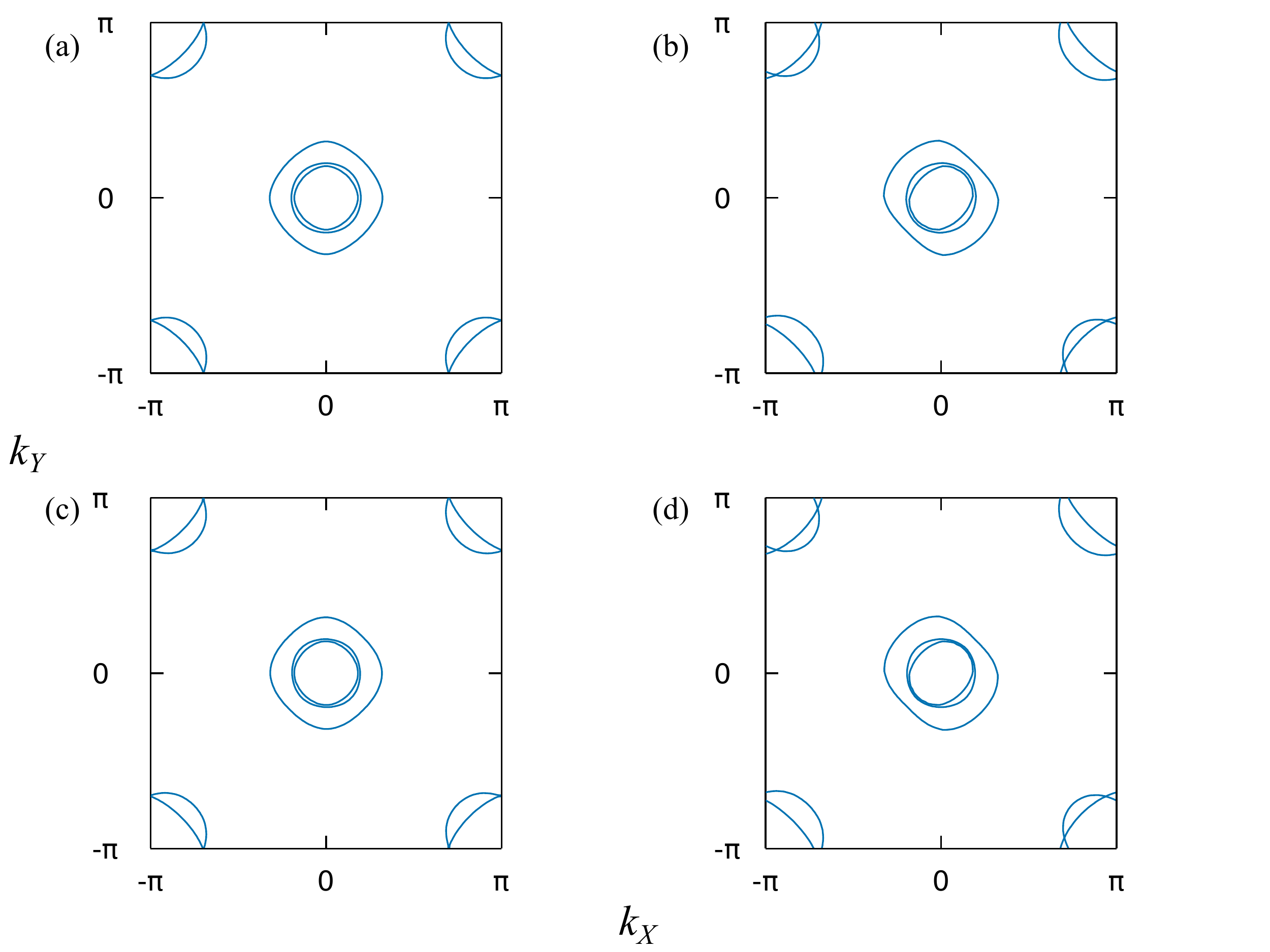}
    \centering
    \caption{Fermi surfaces of tetragonal LaFeAsO [(a)-(b)]
    and orthorhombic LaFeAsO [(c)-(d)]. 
    (a) and (c) the normal state at $T=0.1$, where $\Delta(T)=0$.
    (b) and (d) the orbital ordered state at $T = 0.01$.}
    \label{fig:LaFeAsO_FS}
\end{figure}

We examine two crystal structures.  
One is the tetragonal crystal with the space group $P4/nmm$ and the other is the orthorhombic crystal
with the space group $Cmma$.
The former is realized in the high-temperature phase of LaFeAsO, while the latter emerges below
the structural transition temperature
$T_s \approx 160$ K.
Lattice parameters are derived from  Ref.~\onlinecite{LaFeAsO_Nomura}.
Although the structural transition is associated with the nematic order, we independently study the two phenomena to clarify the origin of the EQMs. 
The Fermi surfaces in the orbital-ordered state (normal state) on the tetragonal crystal lattice are shown in Fig.~\ref{fig:LaFeAsO_FS}(b) (Fig.~\ref{fig:LaFeAsO_FS}(a)), indicating sizable distortion of the Fermi surfaces due to the orbital order. 
The same plots for the orthorhombic crystal lattice are shown in Figs.~\ref{fig:LaFeAsO_FS}(c) and \ref{fig:LaFeAsO_FS}(d). 
We see that the effects of the orthorhombic crystal distortion on the Fermi surfaces 
are not prominent.

\begin{figure}[htbp]
    \centering
    \includegraphics[width=0.8\linewidth]{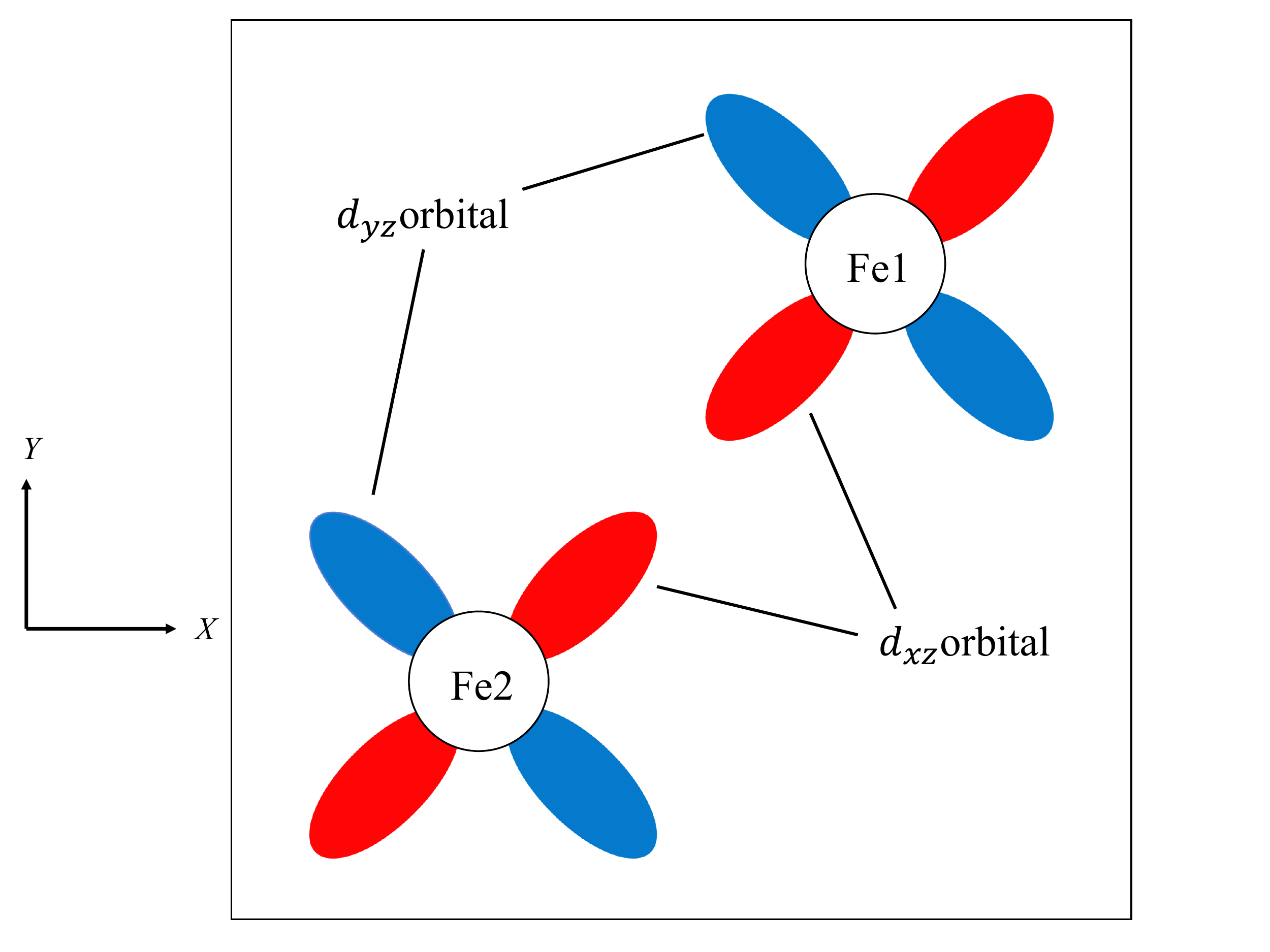}
    \caption{Unit cell of LaFeAsO and FeSe. The $d_{xz}$ and $d_{yz}$ orbitals are illustrated. 
    The $x$-axis is different from the principal $X$ axis of the crystal by $45^\circ$.}
    \label{fig:unit_cell_LaFeAsO}
\end{figure}

We take the unit of energy so that the largest hopping integral of LaFeAsO is $t_{d_{xz(yz)} d_{xz(yz)} 12}=1$, 
which is the nearest neighbour hopping integral between parallel $d_{xz(yz)}$ orbitals. 
The other hopping integrals and temperature are scaled as $t/t_{d_{xz(yz)} d_{xz(yz)} 12}$.
The unit cell and the shape of the two orbitals are illustrated in Fig.~\ref{fig:unit_cell_LaFeAsO}.
Although we may expect a larger hopping integral of red orbitals,  
for the 10-orbital models of $3d$ electrons on Fe ions, 
the hopping between the blue orbitals is larger than that between the red orbitals, 
since the hopping through $2p$ orbitals of As ions is dominant.
Thus, we choose this hopping integral as the unit of energy.
We take the same unit also for FeSe, because
the hopping parameter is not significantly different between the iron-based superconductors. 
Note that the axes for orbitals in Fig.~\ref{fig:unit_cell_LaFeAsO} are rotated by $45^\circ$ from the principal $X$ and $Y$ axes in accordance with the conventional notation.
We adopt the $(x,y)$ axes and corresponding wave vector $(k_x,k_y)$ for calculating the thermodynamic EQMs, although the Fermi surfaces are drawn by using the $(k_X,k_Y)$ axes. In this notation a finite EQM $Q_{x^2-y^2}$ is induced by the orbital order of $d_{xz}$ and $d_{yz}$ orbitals.

\begin{figure}[htbp]
    \centering
    \includegraphics[width=0.9\linewidth]
    {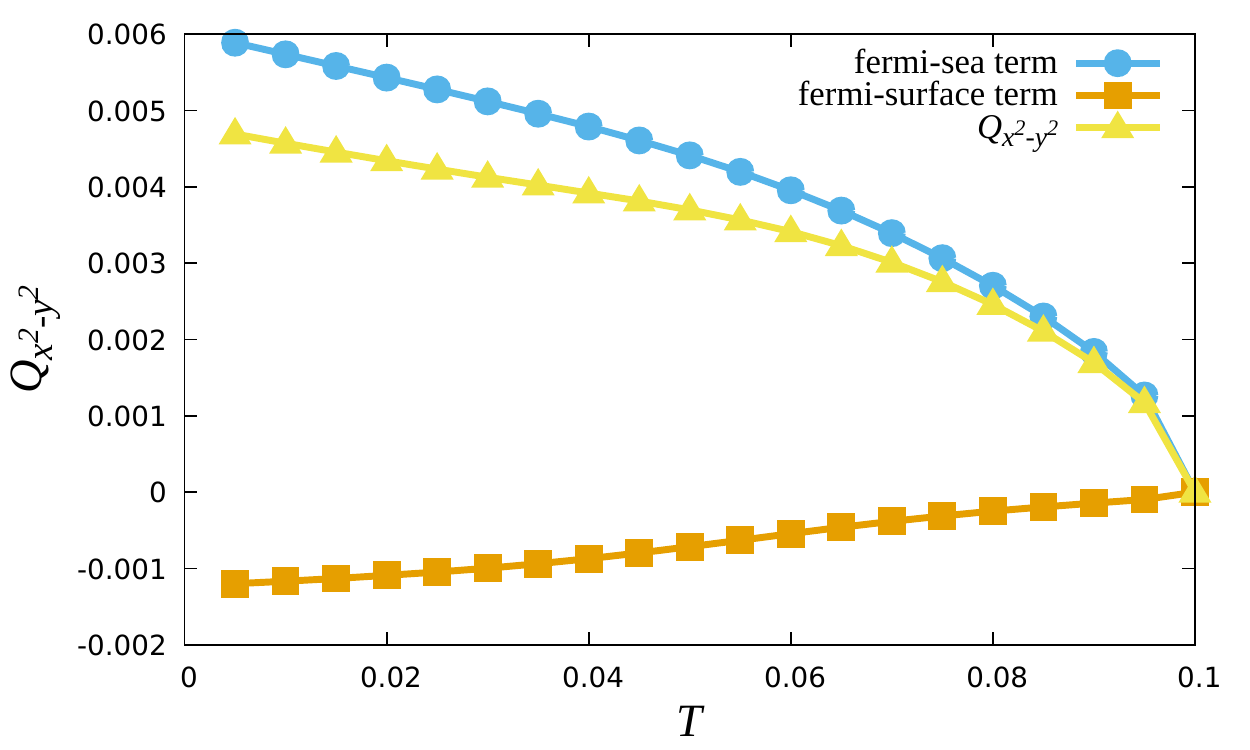}
    \caption{Temperature dependence of the thermodynamic EQM $Q_{x^2-y^2}$ in the tetragonal LaFeAsO.
    We show the contributions from the Fermi-sea term and the Fermi-surface term by circles with blue line and squares with red line, respectively.
    The total EQM is shown by triangles with yellow line.}
    \label{fig:EQM_LaFeAsO_tetra}
\end{figure}
\begin{figure}[htbp]
    \centering
    \includegraphics[width=0.9\linewidth]
    {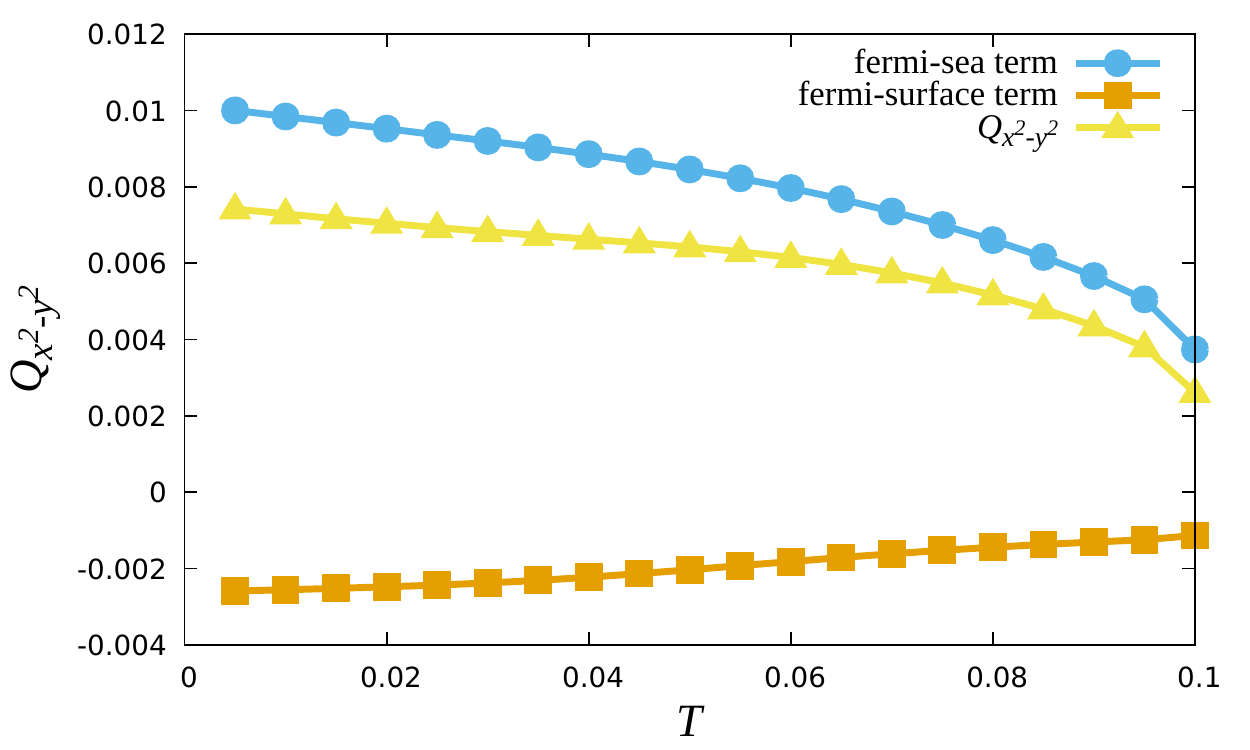}
    \caption{Thermodynamic EQM $Q_{x^2-y^2}$ in the orthorhombic LaFeAsO.}
    \label{fig:EQM_LaFeAsO_ortho}
\end{figure}

The temperature dependences of the thermodynamic EQMs in 
the tetragonal and orthorhombic LaFeAsO are shown in
Fig.~\ref{fig:EQM_LaFeAsO_tetra} and Fig.~\ref{fig:EQM_LaFeAsO_ortho}, respectively.
We show the contributions from the Fermi-sea term and the Fermi-surface term by blue lines and red lines, respectively. Total thermodynamic EQMs are shown by yellow lines. 
The quantities are indicated by the same color and symbols in all later results.

In both tetragonal and orthorhombic structures, the Fermi-sea term is dominant for the EQMs.
Thus, the EQM in the nematic phase of LaFeAsO mainly has a geometric origin. 
Comparing the results for the tetragonal and orthorhombic crystals, we notice the additional contribution to the thermodynamic EQMs from the orthorhombic crystal deformation. We see sizable EQMs at $T = 0.1$ in Fig.~\ref{fig:EQM_LaFeAsO_ortho}, where  $\Delta(T) = 0$.

\begin{figure}[tbp]
    \centering
    \includegraphics[width=0.9\linewidth]{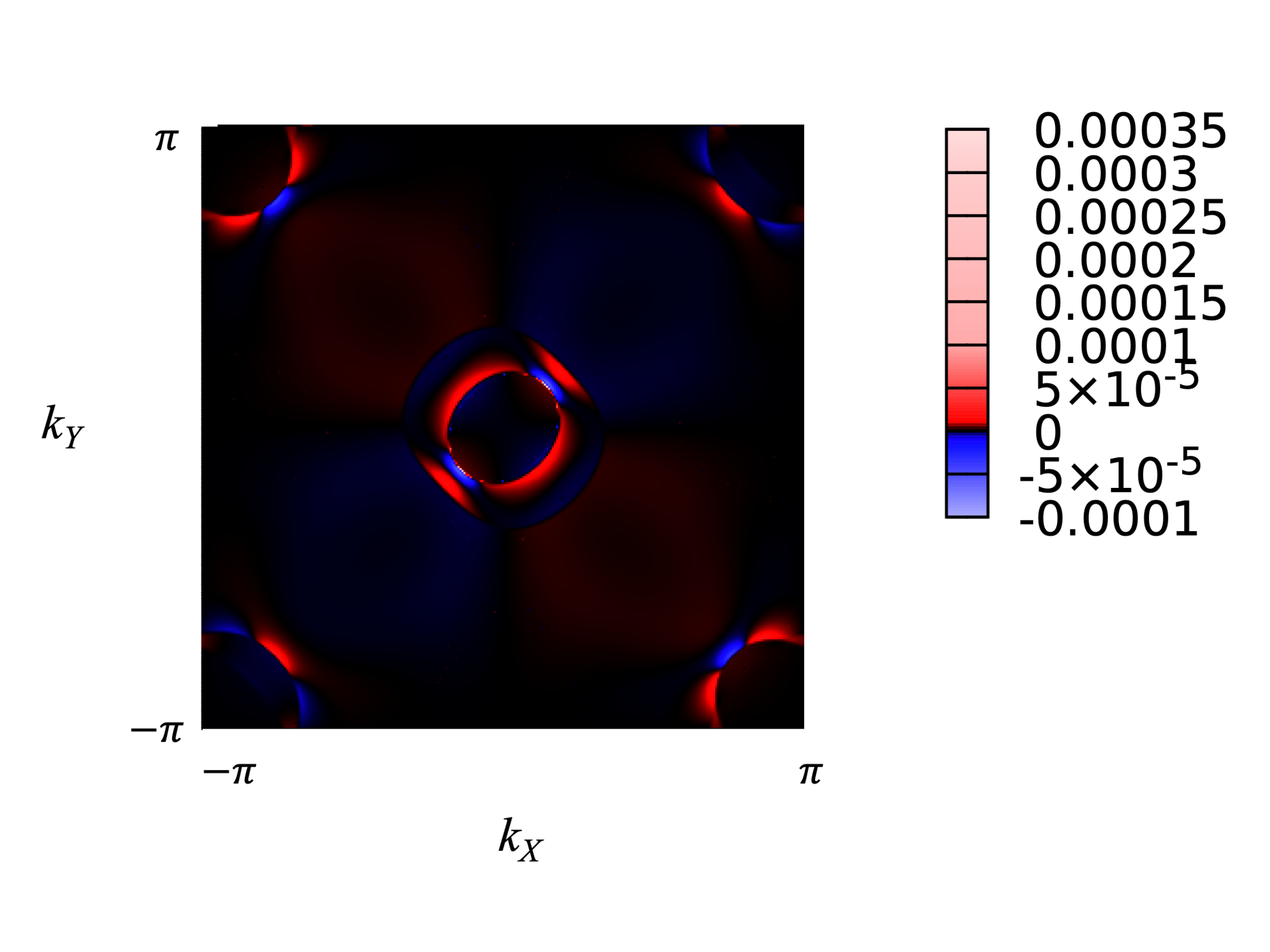}
    \caption{Fermi-sea term of the tetragonal LaFeAsO from each $\bm k$ points. We set $T = 0.01$.
    Large contributions from the momentum near the Fermi surfaces around the $\Gamma=(0,0)$ and $M=(\pi,\pi)$ points are observed.}
    \label{fig:LaFeAsO_sea_kdep0.01}
\end{figure}

\begin{figure}[htbp]
    \centering
    \includegraphics[width=0.45\linewidth]{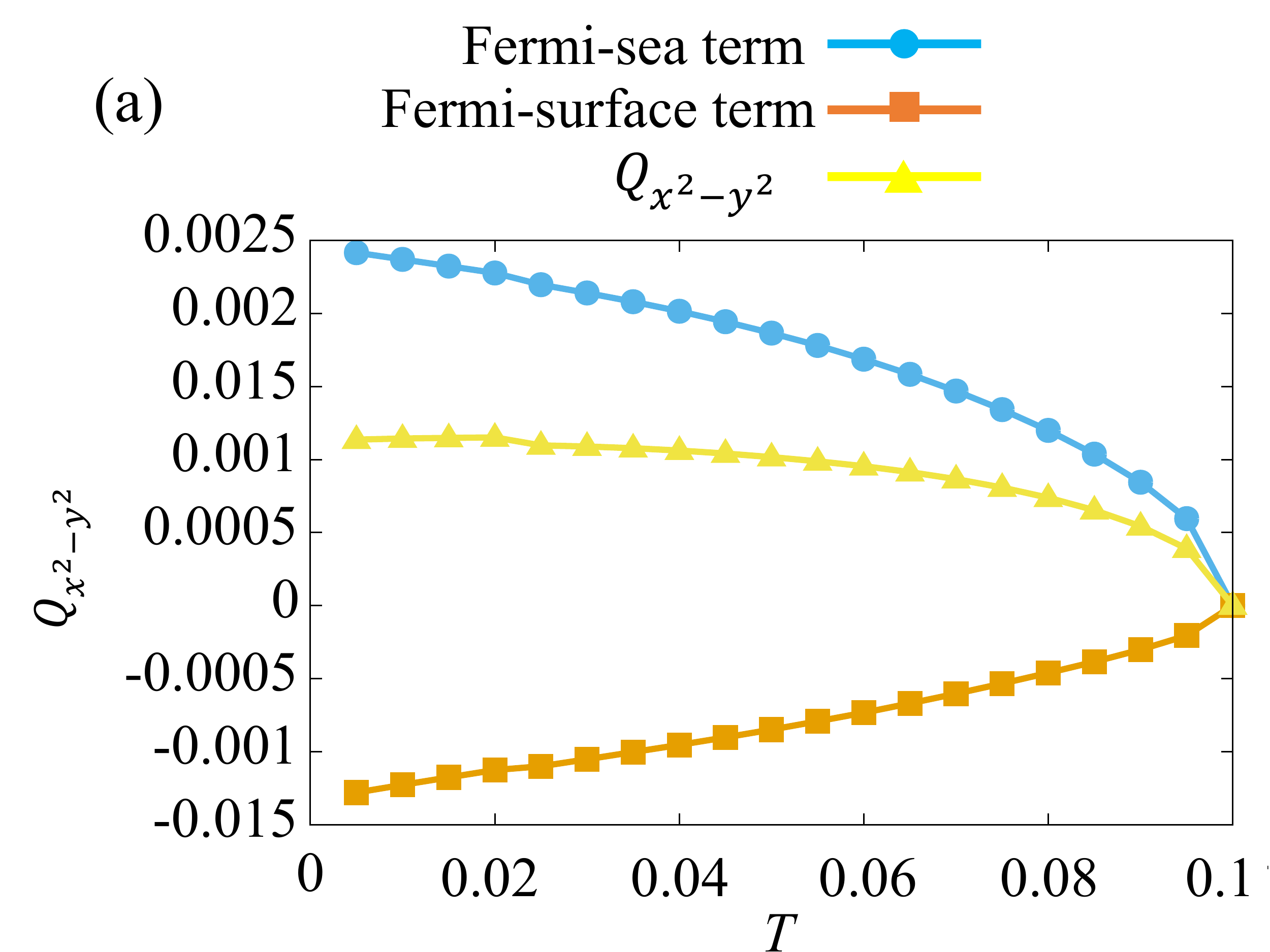}
    \hspace*{5mm}
     \includegraphics[width=0.45\linewidth]{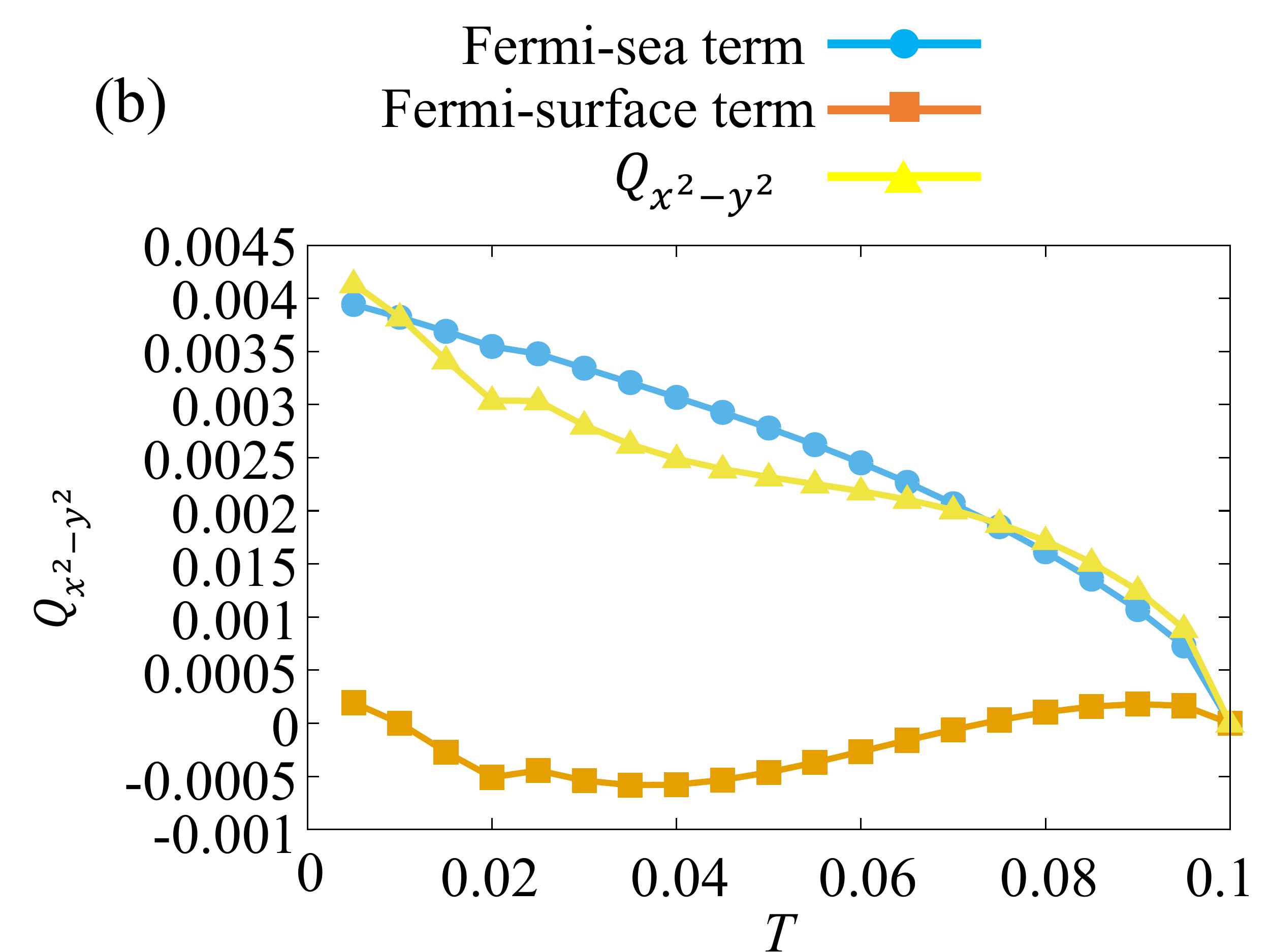}
    \caption{Contributions to the thermodynamic EQM $Q_{x^2-y^2}$ in \YY{LaFeAsO} from momentum spaces (a) $-\pi/3< k_x, k_y < \pi/3$ and (b) $2\pi/3< k_x, k_y < 4\pi/3$.
    Not only the Fermi-surface term but also the Fermi-sea term mainly come from the momentum near the Fermi surfaces (see also Fig.~\ref{fig:LaFeAsO_sea_kdep0.01}). 
}
    \label{fig:LaFeAsO_EQMs_Gamma}
\end{figure}

Here we show that the dominant geometric origin of the EQMs is a unique property of LaFeAsO by comparing the 10-orbital model from first-principles with a toy model.  
Two of us previously calculated the thermodynamic EQMs\cite{TEQMs} using a toy model constructed for only the $d_{xz}$ and $d_{yz}$ orbitals\cite{Raghu}. 
The Fermi-sea term with a geometric origin is much larger in the 10-orbital model than the toy model. 
This is because the Fermi-sea term is enhanced by the band degeneracy. 
To see this we show the momentum-resolved Fermi-sea term in Fig.~\ref{fig:LaFeAsO_sea_kdep0.01}. 
Dominant contributions come from the $k$ points near the Fermi surfaces. 
Thus, in LaFeAsO, the geometrically nontrivial property of wave functions due to the band degeneracy around the Fermi surfaces gives rise to the sizable Fermi-sea term (see also Appendix \ref{appendix:degeneracy}). 
The EQMs from the momentum near the $\Gamma$ and $M$ points are plotted in Fig.~\ref{fig:LaFeAsO_EQMs_Gamma}.
Figures~\ref{fig:LaFeAsO_EQMs_Gamma}(a) and \ref{fig:LaFeAsO_EQMs_Gamma}(b) show the EQM arising from the momentum $|k_x|, |k_y| < \pi/3$ and that from $|k_x -\pi|, |k_y -\pi| < \pi/3$, respectively.
As shown in Figs.~\ref{fig:LaFeAsO_sea_kdep0.01} and \ref{fig:LaFeAsO_EQMs_Gamma}, the main contribution comes from the Fermi surfaces around the $M$ point. 
On the other hand, the degenerate band structure is too simplified in the toy model, and in particular, the contribution from the momentum near the $M$ point is almost overlooked.
While the electron and hole Fermi surfaces are additive for the Fermi-sea term, the Fermi-surface term is partially cancelled. 
Given that the multiple band degeneracy near the Fermi surfaces is a unique property of LaFeAsO, the geometric origin of the EQM is also regarded as a characteristic property of LaFeAsO. 
To support this argument, we calculate the chemical potential dependence of the EQMs in the 10-orbital model and find that the Fermi-surface term is comparable or larger than the Fermi-sea term in most cases except for the realistic parameter of LaFeAsO (see Appendix~\ref{appendix:LaFeAsO_mu_dep}).

Note that the geometrical origin of the EQMs is not a unique consequence of the orbital order. Indeed, the orthorhombic lattice distortion causes not only the orbital polarization but  also the bond anisotropy, and the magnitude of the bond order is larger than that of the orbital order. We see a sizable Fermi-sea term due to the lattice distortion (Fig.~\ref{fig:EQM_LaFeAsO_ortho} at $T=0.1$), which is much larger than the Fermi-surface term.

\subsection{FeSe\label{FeSe}}
Conducting a first-principles calculation for FeSe, we construct the 10-orbital tight-binding model similar to LaFeAsO.
Lattice parameters given in Ref.~\onlinecite{FeSe_Bohmer}
are adopted, and the space group is $P4/nmm$.
It is known that 
in FeSe tiny Fermi-surfaces obtained by
angle-resolved photoemission spectroscopy (ARPES) measurements~\cite{Maletz_ARPES,Zhang_ARPES} are not reproduced by the first-principle calculation: 
larger Fermi surfaces and an extra Fermi surface of the $d_{xy}$ orbital appear. 
To reproduce the experimentally observed Fermi surfaces of FeSe, we take into account additional hopping parameters in addition to those given by the \wien code, in a similar manner to Refs.~\cite{ohnari_sign,Yamakawa2016,Ishizuka2018} (see Appendix.~\ref{appendix:FeSe_FS} for details).
The additional hopping parameters may stem from the self-energy correction \cite{Gorni2021}.
Different from LaFeAsO, sign reversal of the orbital polarization in the momentum space between the $\Gamma$ and $M$ points has been observed by ARPES~\cite{nakayama_sign,Shimojima_sign,Suzuki_sign,watson_sign,zhang_sign,Zhang_ARPES,Maletz_ARPES} for FeSe 
and studied theoretically~\cite{ohnari_sign}.
To reproduce this property of the nematic order, we take into account the molecular fields of the bond order in addition to the orbital order
(see Appendix.~\ref{appendix:FeSe_orbital} for details).
The total molecular field is given by
\begin{eqnarray}\label{eq:FeSe_Hamiltonian}
    &&\Gamma = \Gamma_{\rm orb} + \Gamma_{\rm bond},\\
    &&\Gamma_{\rm orb} =
    \sum_{\bm k} \sum_{\sigma, i=1,2}
    \left[
    c^{\dagger}_{d_{xz}i\sigma}(\bm k)c_{d_{xz}i\sigma}(\bm k)
    -
    c^{\dagger}_{d_{yz}i\sigma}(\bm k)c_{d_{yz}i\sigma}(\bm k)\right],\notag\\
    \label{eq:FeSe_orbital} \\
    &&\Gamma_{\rm bond} = 
    \sum_{\bm k} 2\left(
    \cos\frac{k_y-k_x}{2}-\cos\frac{k_y+k_x}{2}
    \right)\notag\\
    &&\ 
    \times\sum_{\sigma, l = xz,yz}\left[
    c^\dagger_{d_l 1 \sigma}(\bm k)c_{d_l 2 \sigma}(\bm k)
    +c^\dagger_{d_l 2 \sigma}(\bm k)c_{d_l 1 \sigma}(\bm k)
    \right].\label{eq:FeSe_bond}
\end{eqnarray}

\begin{figure}[tbp]
    \centering
    \includegraphics[width=1.0\linewidth]
    {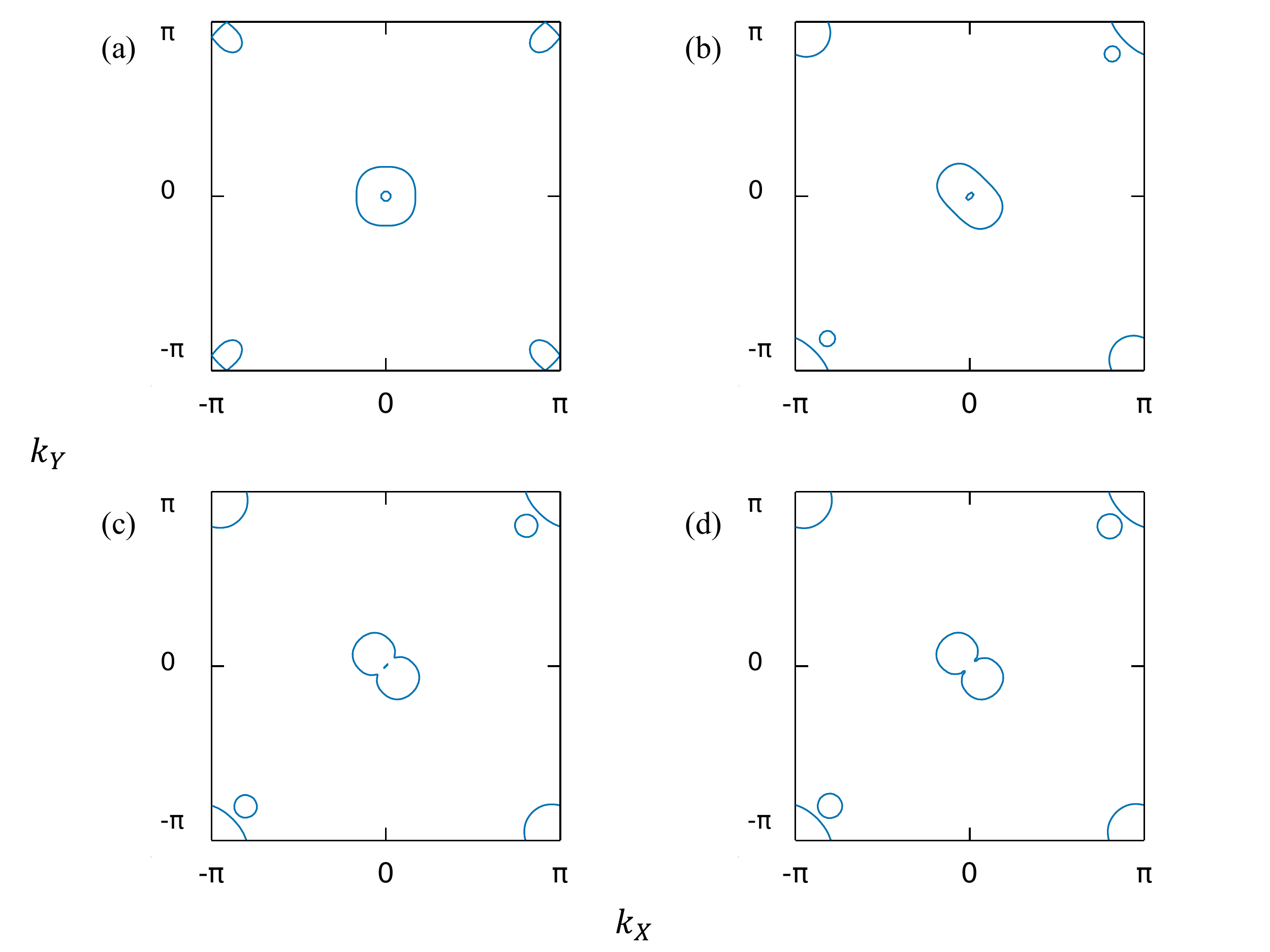}
    \caption{Fermi surfaces of FeSe at
    (a) $T = 0.1$,
    (b) $T = 0.05$,
    (c) $T = 0.02$, and 
    (d) $T = 0.01$.
    (a) is the normal state, while (b)-(d) are the nematic states with finite $\Delta(T)$.
    }
    \label{fig:FeSe_FS_all}
\end{figure}

The Fermi surfaces of the 10-orbital model with additional hopping parameters and nematic order parameter $\Delta(T)$
are shown in Fig.~\ref{fig:FeSe_FS_all}.
The Fermi surfaces are distorted with growing the nematic order. 
In the low temperature region [Fig.~\ref{fig:FeSe_FS_all}(d)], one Fermi surface near the $\Gamma$ point disappears owing to the orbital polarization, consistent with experiments~\cite{Shibauchi2020}. 
The disappearance of the Fermi surface is related to the change in the Fermi-sea term of the EQM that will be shown below.
Note that the shape of the remaining Fermi surfaces in Fig.~\ref{fig:FeSe_FS_all}(d) is slightly different from what observed in the experiment since we do not take into account a weak spin-orbit coupling~\cite{ohnari_sign}.
For a remark, we need to calculate the chemical potential at each temperature $T$ to keep the particle number and reproduce the disappearance of the Fermi surface. 

\begin{figure}[htbp]
    \centering
    \includegraphics[width=0.9\linewidth]
    {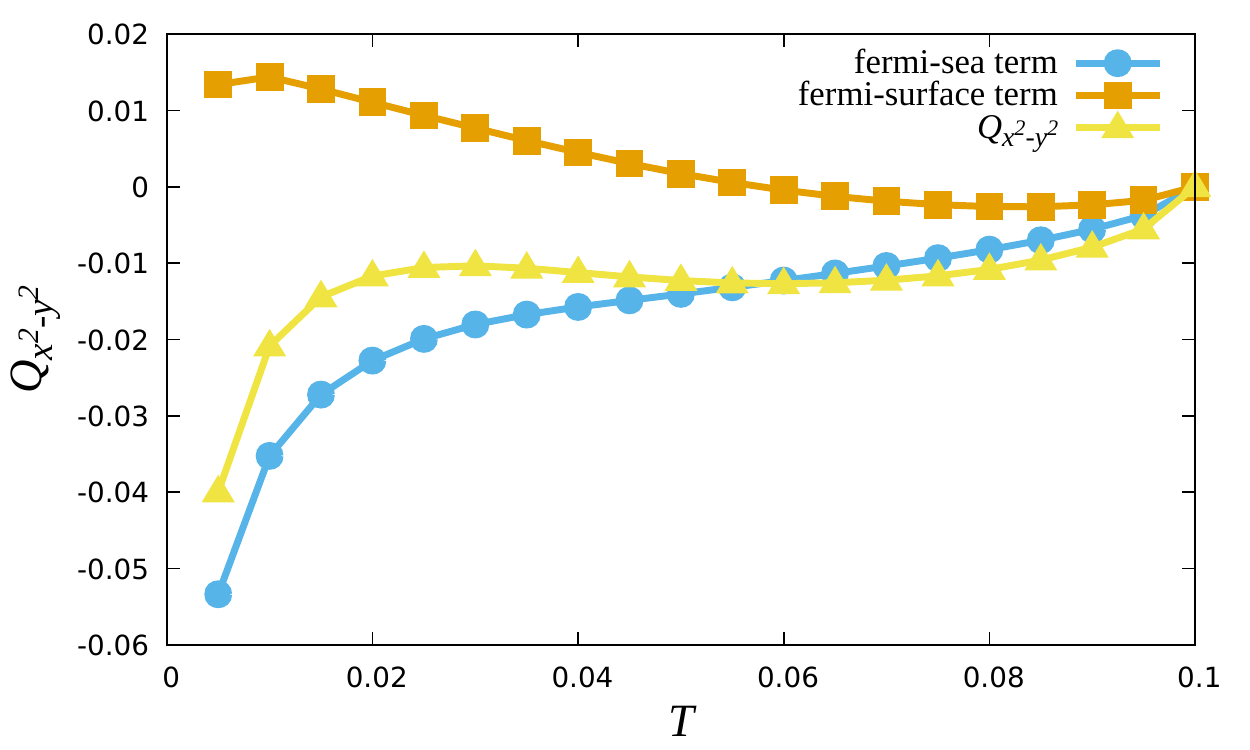}
    \caption{Thermodynamic EQM $Q_{x^2-y^2}$ in FeSe.
}
    \label{fig:FeSe_EQMs}
\end{figure}

\begin{figure}[htbp]
    \centering
    \includegraphics[width=0.45\linewidth]{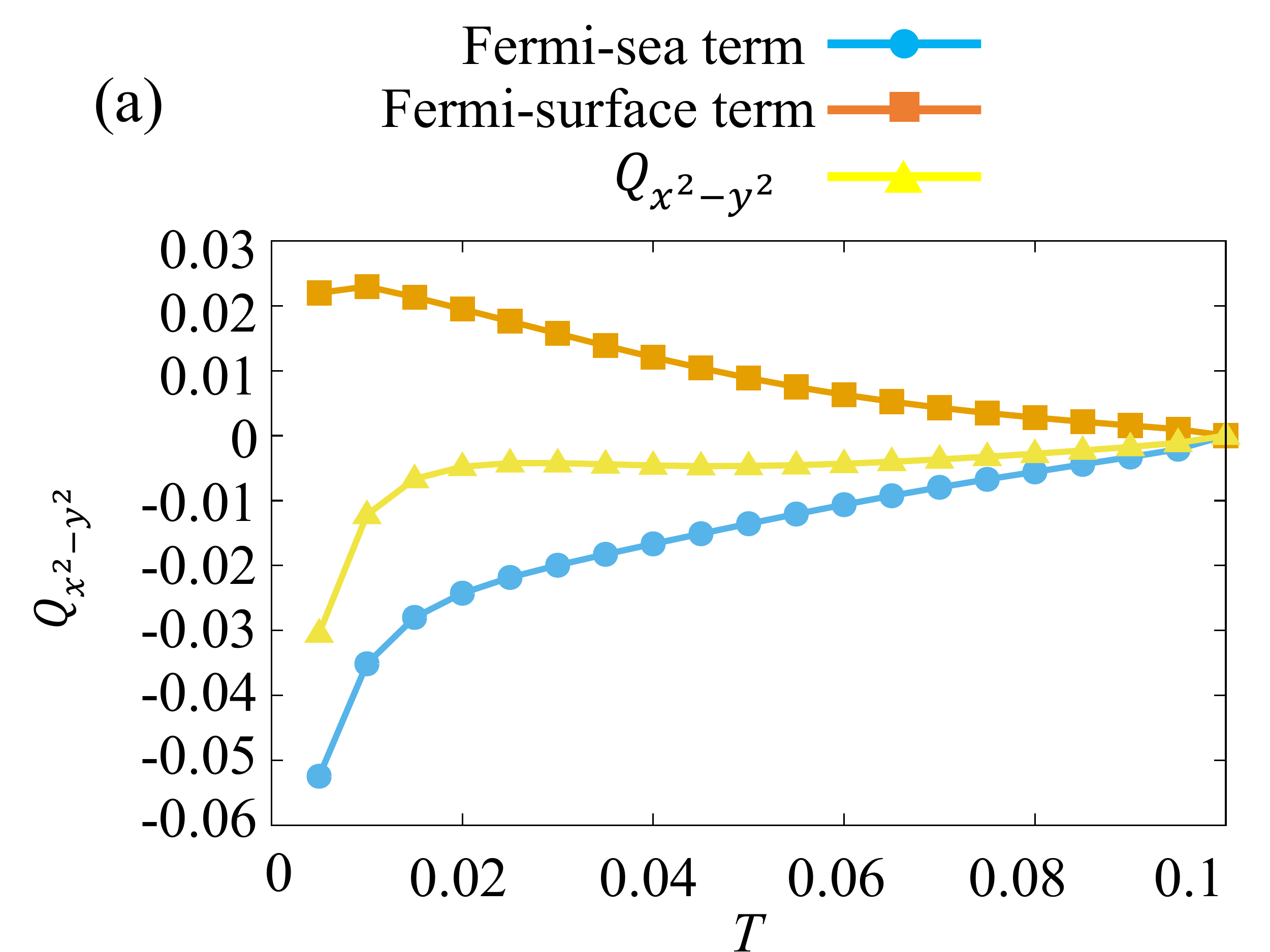}
    \hspace*{5mm}
     \includegraphics[width=0.45\linewidth]{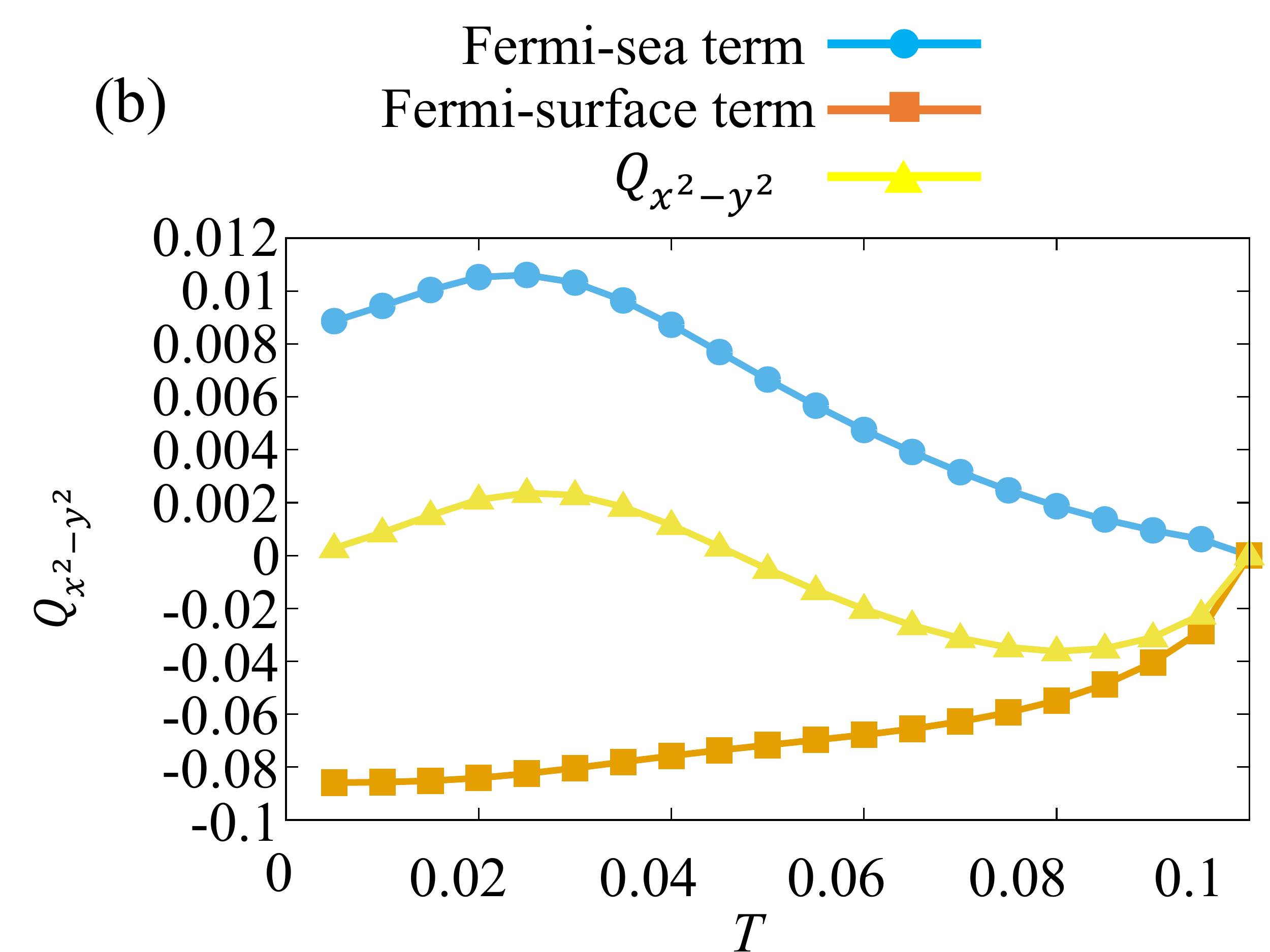}
    \caption{Contributions to the thermodynamic EQM $Q_{x^2-y^2}$ in FeSe from (a) $-\pi/3< k_x, k_y < \pi/3$ and (b) $2\pi/3< k_x, k_y < 4\pi/3$.
    Both Fermi-sea and Fermi-surface terms mainly originate from the momentum near the Fermi surfaces (see also Fig.~\ref{fig:FeSe_sea_kdep}). 
}
    \label{fig:FeSe_EQMs_Gamma}
\end{figure}

The EQM $Q_{x^2-y^2}$ in FeSe is shown in Fig.~\ref{fig:FeSe_EQMs}. 
At low temperatures, the Fermi-sea term with a geometric origin is also dominant in FeSe.
In contrast to LaFeAsO, the Fermi-sea term is negative.
This contribution mainly comes from the electronic states near the $\Gamma$ point as we show Fig.~\ref{fig:FeSe_EQMs_Gamma}.
We see the negative and dominant contribution to the Fermi-sea term from near the $\Gamma$ point and the positive contribution from near the $M$ point.
As for the Fermi-surface term, Fig.~\ref{fig:FeSe_EQMs_Gamma} also shows that the dominant contribution comes from near the $\Gamma$ point. 

\begin{figure}[htbp]
    \centering
    \includegraphics[width=0.8\linewidth]
    {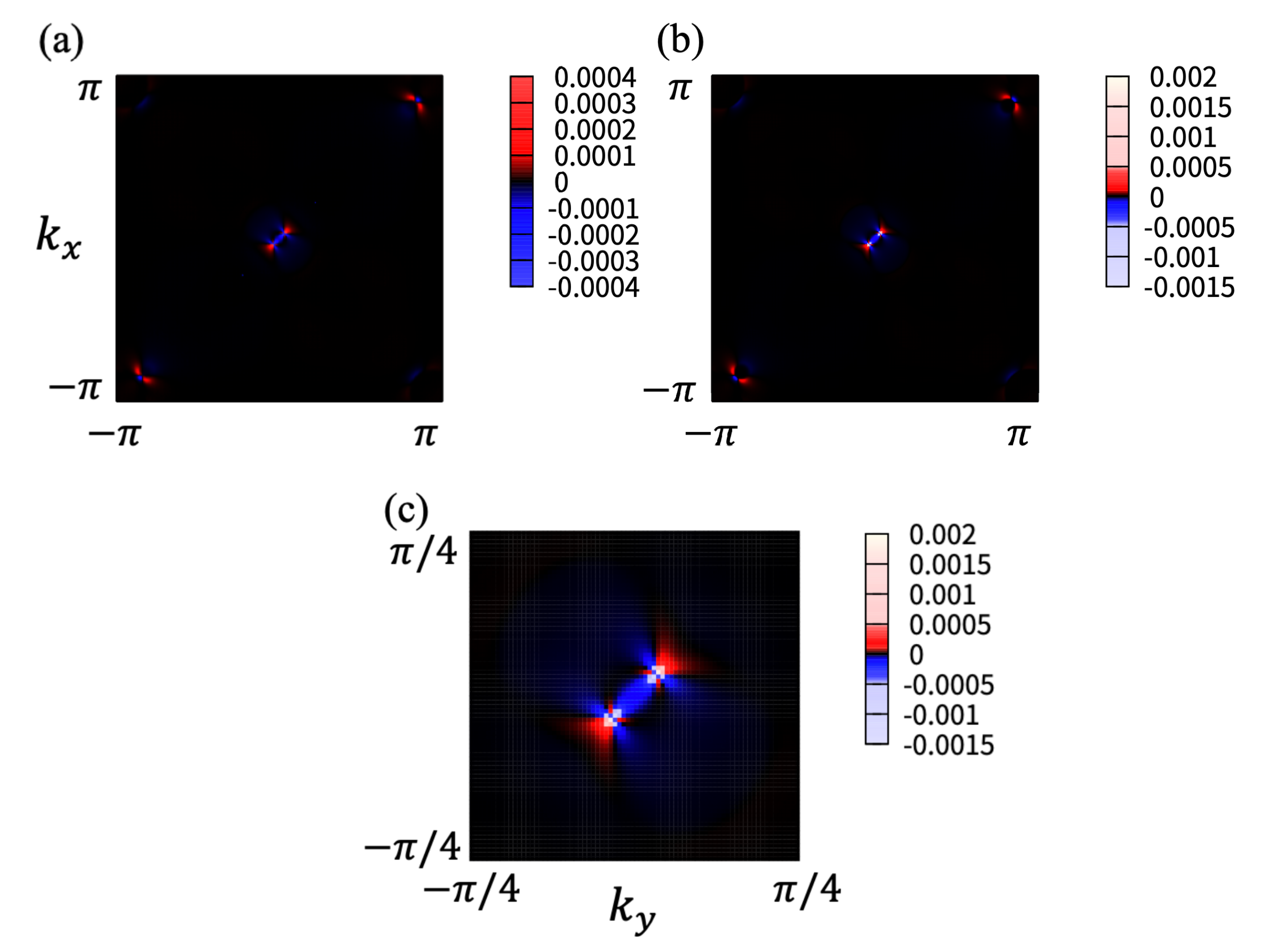}
    \caption{Fermi-sea term of the EQM in FeSe from each $\bm k$ points for (a) $T = 0.02$ and (b) $T=0.01$.
    (c) shows an enlarged illustration of (b) near the $\Gamma$ point.
    The contribution from near the Fermi surfaces dominates the Fermi-sea term.}
    \label{fig:FeSe_sea_kdep}
\end{figure}

For more details, the Fermi-sea term contributions from each $\bm k$ points at $T = 0.02$ and $T = 0.01$ are shown in Figs.~\ref{fig:FeSe_sea_kdep}(a) and \ref{fig:FeSe_sea_kdep}(b,c), respectively. 
The opposite contribution from the $\Gamma$ and $M$ points is revealed, consistent with Fig.~\ref{fig:FeSe_EQMs_Gamma}.
Furthermore, we see a change in the Fermi-sea term arising from near the $\Gamma$ point between $T = 0.02$ and $T=0.01$, while that from the $M$ point is almost temperature independent in this region. 
This change is caused by the disappearance of a Fermi surface discussed above.
In Fig.~\ref{fig:FeSe_sea_kdep}(c), we see a large contribution, which is illustrated by white color, 
from the momentum around which the Lifshitz transition occurs.

Finally, we discuss the similarities and differences between LaFeAsO and FeSe. 
From the results, we find that the Fermi-sea term with geometric origin is dominant in the thermodynamic EQM of FeSe as well as of LaFeAsO. 
This finding implies that the geometrically nontrivial properties of wave functions
are ubiquitous in iron-based superconductors. Because the geometric properties are owing to the multi-orbital and multi-band structure, the band degeneracy naturally plays important roles for the EQM as well as for the nematic order and superconductivity. 
On the other hand, when we look at the details, the sign of the EQM is opposite between FeSe and LaFeAsO, and the momentum-resolved EQM shows different structures. 
Thus we need a precise model taking account of the realistic electronic structure for quantifying the EQMs of nematic phases.

We would like to stress the usefulness of the thermodynamic formulation for the EQMs. By the thermodynamic EQMs, the nematic order can be quantified in a unified way, even when not only the electronic structures but also the nematic order parameters are different between the materials as in the cases of LaFeAsO and FeSe.

\subsection{Cuprate superconductors\label{teqms_cuprate}}

To illuminate the unique properties of iron-based superconductors, that is, multi-band structure and resulting geometric origin of the EQM, 
we here calculate the thermodynamic EQM of cuprate superconductors for a comparison.
For the nematic order in cuprate superconductors, we consider the $d_{x^2-y^2}$-wave bond order studied extensively
\cite{Yamase2000,Halboth2000,Honerkamp2001,Metzner2012,bulut2013spatially,Kee2004,wang2014charge,sachdev2013bond,berg2009striped,yamakawa2015spin,kawaguchi2017competing,tsuchiizu2018multistage}.
For a comparison, the orbital order of O2$p_x$ and O2$p_y$ orbitals
is also studied later.
Evaluation of translation-symmetry-breaking order, such as the CDW and PDW order, is left for future studies.

For the study of cuprate superconductors, 
we construct the 17-orbital
tight-binding model,
which consists of
the $3d$ orbitals of coppers and
the $2p$ orbitals of oxygens in a unit cell,
using the \wien and Wannier90 code. 
The space group is $I4/mmm$
and lattice parameters
are adopted from Ref.~\onlinecite{1987_La2CuO4}
at $T = 295$K.
The tight-binding parameters are derived for the representative mother compound La$_2$CuO$_4$.
Although La$_2$CuO$_4$ is a Mott insulator and the nematic order occurs by hole doping~\cite{sato2017thermodynamic,Daou2010}, we adopt the half-filling model with $n = 5.0$, since the dependence on the carrier density is negligible.
The model takes into account four oxygen ions and one copper ion in the unit cell. Thus, creation operators of the $2p$ electrons have index for the sublattice. The two oxygens are located on the CuO$_2$ plane, while the other two are apical oxygens.
We set the unit of energy so that the largest nearest-neighbor $d$-$p$ hopping $t_{d_{x^2-y^2}p_{x}}$ is unity.

\begin{figure}[htbp]
    \centering
    \includegraphics[width=1.0\linewidth]{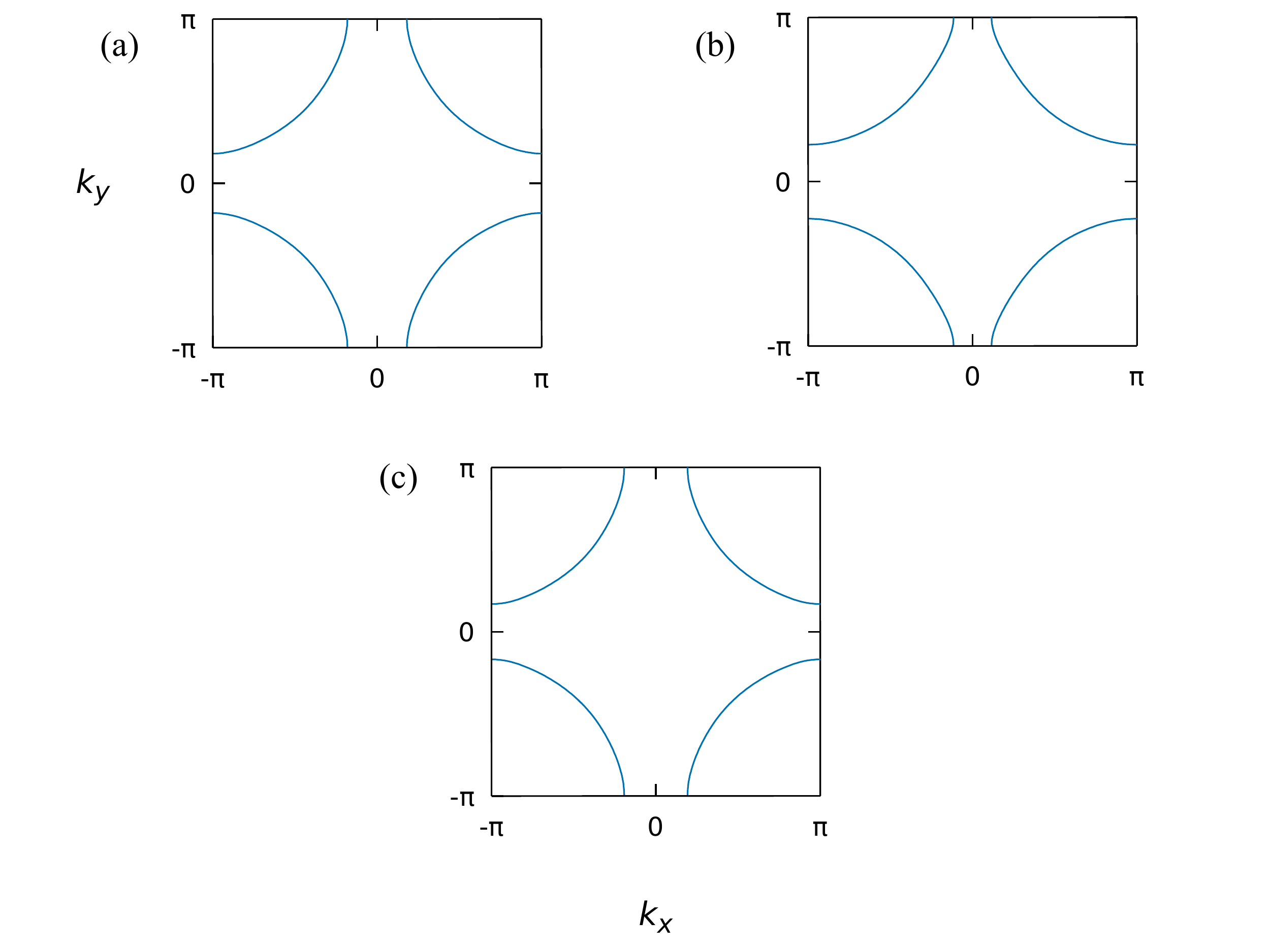}
    \caption{Fermi surface of the $17$-orbital model for La$_2$CuO$_4$. 
    (a) The normal state at $T=0.1$ [$\Delta(T)=0$].
    (b) The $d_{x^2-y^2}$-wave bond-ordered state at $T = 0.01$.
    (c) The $p$-orbital-ordered state at $T = 0.01$.}
    \label{fig:La2CuO4_fs}
\end{figure}

As for the nematic order parameter, the molecular field of the $d_{x^2-y^2}$-wave bond order is given by
\begin{align}
    \Gamma =  \sum_{\bm k} \sum_{\sigma} \left(\cos k_x-\cos k_y\right)
    c^{\dagger}_{d_{x^2-y^2} \sigma}(\bm k) c_{d_{x^2-y^2}\sigma }(\bm k). 
    \label{eq:Hamiltonian_cuprate_bond}
\end{align}
For later comparison, we also examine the $p$-orbital order whose molecular field is written as
\begin{align}
    \Gamma = \sum_{\bm k} \sum_{i=1,2}
    \left[
    c^{\dagger}_{p_{x}i\sigma}(\bm k)c_{p_{x}i\sigma}(\bm k)
    -c^{\dagger}_{p_{y}i\sigma}(\bm k)c_{p_{y}i\sigma}(\bm k)
    \right].
\end{align}
The index $i=1,2$ indicates the oxygens on the CuO$_2$ plane.
Figure~\ref{fig:La2CuO4_fs} shows distortion of the Fermi surface due to the nematic order. It is significant in the bond-ordered state, because the electronic states near the Fermi level mainly consist of the $d_{x^2-y^2}$ orbital, although it is hybridyzed with the $p$ orbitals.

\begin{figure}[htbp]
    \centering
    \includegraphics[width=0.8\linewidth]
    {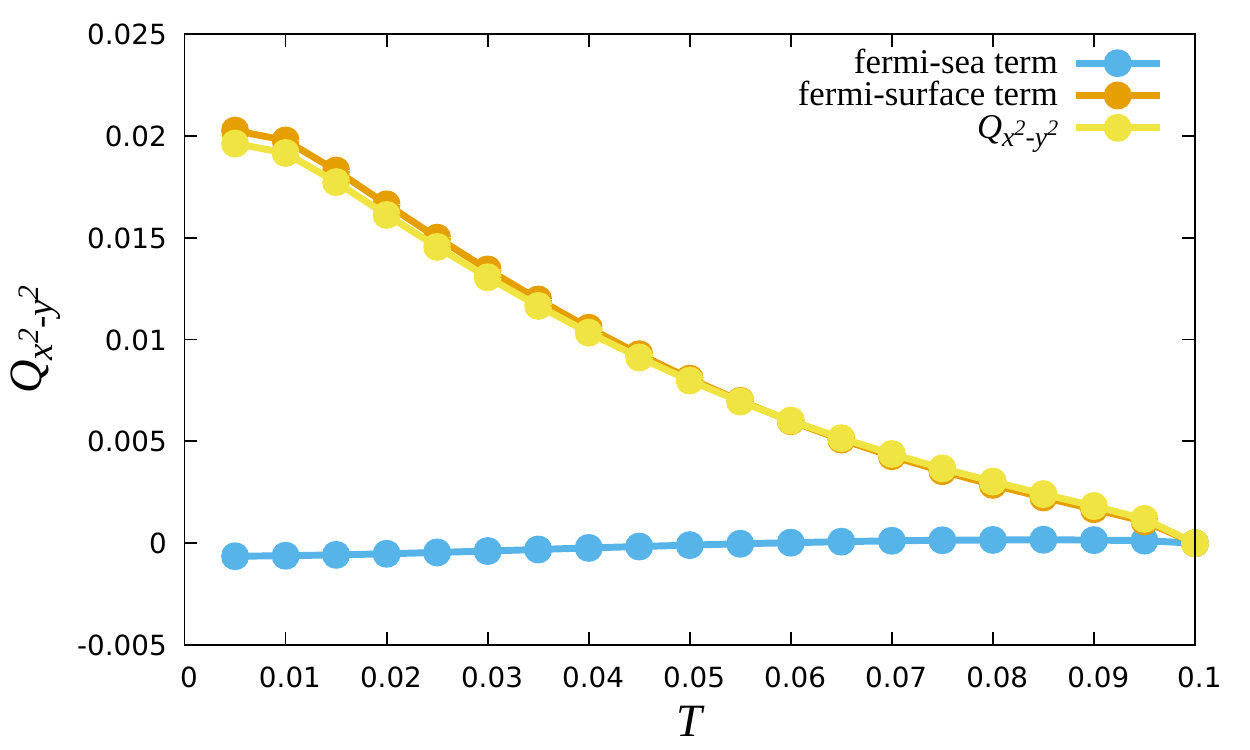}
    \caption{Thermodynamic EQM in the $17$-orbital $d$-$p$ model for La$_2$CuO$_4$ with the $d_{x^2-y^2}$-wave bond order.
}
    \label{fig:La2CuO4_17_bond_EQMs}
\end{figure}

\begin{figure}[htbp]
    \centering
    \includegraphics[width=0.8\linewidth]
    {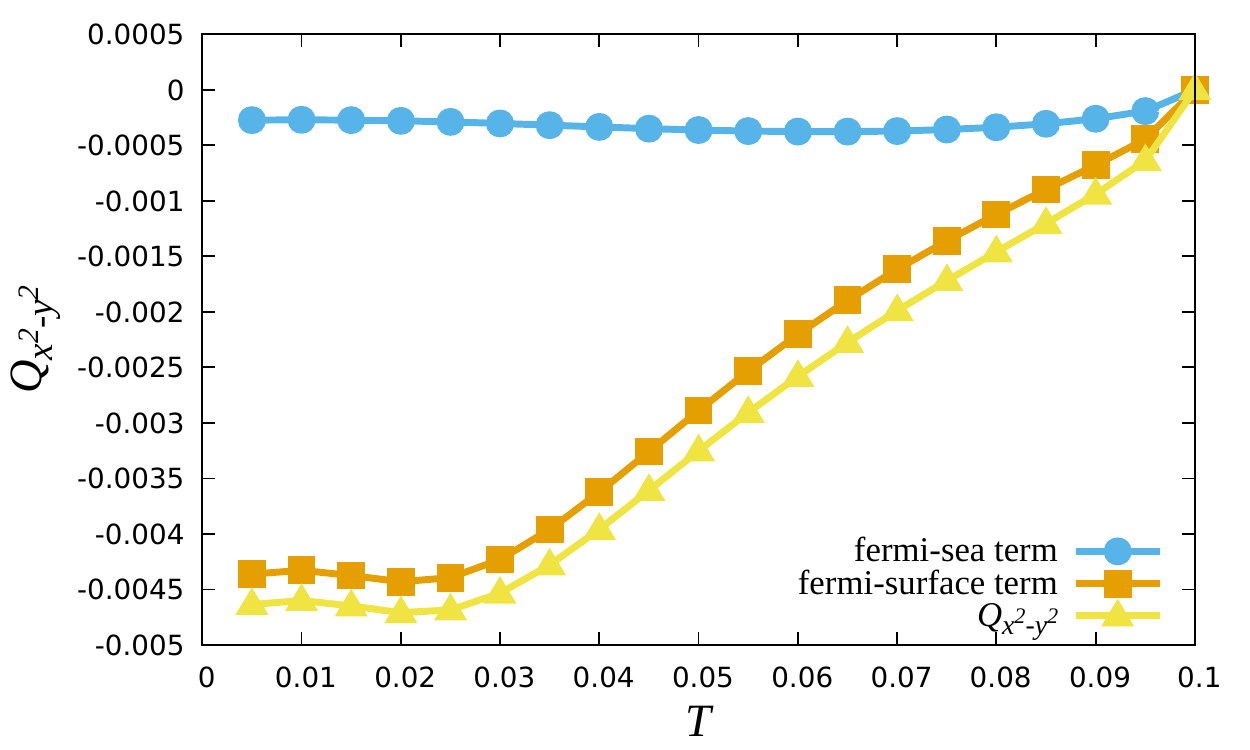}
    \caption{Thermodynamic EQM in the $17$-orbital $d$-$p$ model for La$_2$CuO$_4$ with the orbital order.
}
    \label{fig:La2CuO4_17_orbital_EQMs}
\end{figure}

The thermodynamic EQM induced by the bond order is shown in Fig.~\ref{fig:La2CuO4_17_bond_EQMs}. 
We see that the Fermi-surface term is dominant in contrast to the results for the iron-based superconductors.
Unlike the iron-based superconductors, the band near the Fermi level is isolated from others, although the hybridized $d$-$p$ orbital forms the Fermi surface. 
Comparison between the iron-based and cuprate superconductors implies a unique property of the former from the viewpoint of the EQM; the band degeneracy near the Fermi level gives rise to geometrically nontrivial properties that lead to the dominant Fermi-sea term of the EQM. 
We would like to stress that the difference mainly comes from the underlying electronic structure and not from the character of nematic order parameters. Indeed, the orbital order also induces the dominant Fermi-surface term in cuprate superconductors, as we see in Fig.~\ref{fig:La2CuO4_17_orbital_EQMs}.

In Appendix~\ref{appendix:dp-model}, we show the qualitatively same results for the three-orbital $d$-$p$ model, which has been extensively analyzed in the previous studies of cuprate superconductors. 
Thus, just increasing the number of orbitals does not enhance the Fermi-sea term. The band degeneracy near the Fermi surface is an essential condition for a large Fermi-sea term. 
Comparison between the $d_{x^2-y^2}$-wave bond order and the $p$-orbital order shows a larger EQM in the former, although we assume the same energy scale of the order parameters. 
This is simply because the Fermi-surface term is dominant in both cases, and because the distortion of the Fermi surfaces is small in the $p$-orbital-ordered state.

\section{summary and discussion}
In this paper,
after showing the failure of the EQMs given by the electromagnetism and Wannier function methods, 
we evaluated the thermodynamic EQMs in LaFeAsO, FeSe, and La$_2$CuO$_4$ using the first-principles calculation and assuming the candidate nematic order parameters. 
The thermodynamic EQMs have been proposed as one of the fundamental quantities characterizing the $C_4$-symmetry breaking in the various nematic phases. 
From the results,
we found that the EQMs in iron-based superconductors have a geometric origin. 
This is due to the highly degenerate band structure near the Fermi level, unique to iron-based superconductors. 
In contrast, the EQMs of cuprate superconductors mainly originate from the distortion of Fermi surfaces. In this case, the magnitude and sign of the EQMs can be derived from the band structure, which can be observed by ARPES for instance: The third term of Eq.~\eqref{eq:TEQMs} gives the EQMs. Thus, differences between iron-based superconductors and cuprate superconductors in the nematic phases were elucidated from the perspective of the EQMs. 

In addition to the conceptual meaning characterizing the $C_4$-symmetry breaking, the EQMs are related to some electric responses caused by the symmetry breaking. 
A thermodynamic relation between the EQMs and the electric susceptibility~\cite{TEQMs} 
\begin{align}
\frac{\partial Q_{ij}(\mu)}{\partial \mu} = -\chi^e_{ij}, 
\end{align}
has been proved for insulators at $T=0$. 
This relation implies that the geometric contribution plays an essential role for the electric susceptibility in the insulating ground state of iron-based superconductors' mother compounds.

We also see indirect relations of the EQMs with some optical responses. 
For example, the optical attenuation coefficient is given by~\cite{sipe2000second-order,iba2018ab,nastos2008full},
\begin{eqnarray}
    \varepsilon_{\rm att}^{ij}(\omega) &=& i\pi \sum_{n\neq m}\int\dfrac{d^dk}{(2\pi)^d}g_{nm}^{ij}(\bm k) \left[f(\epsilon_n(\bm k))-f(\epsilon_m(\bm k))\right]\notag\\
    &&\times \delta(\epsilon_m(\bm k) - \epsilon_n(\bm k)-\hbar\omega). 
\end{eqnarray}
Here, $g_{nm}^{ij}(\bm k) = \frac{1}{2}(A_{nm}^i(\bm k) A_{mn}^j(\bm k) + c.c)$ is the band-resolved quantum metric, which also appeared in the thermodynamic EQMs. The anisotropic optical attenuation caused by the nematic $C_4$-symmetry breaking may be related to the geometric term of the thermodynamic EQMs. 
As for the relation to the nonlinear optics, the photocurrent responses in time-reversal-symmetric and $PT$-symmetric systems have been recently classified, and the results reveal that the quantum metric is an essential quantity for some photocurrent responses, such as shift current, magnetic injection current, and gyration current~\cite{watanabe2021chiral,Ahn2020}.
Thus, elucidation of linear and nonlinear optical responses in iron-based superconductors may be an intriguing future issue.
For the photocurrent generation, the space inversion symmetry must be broken. Indeed, the inversion symmetry is broken in some iron-based superconductors, such as FeSe/SrTiO$_3$~\cite{Wang_2012,Liu2012,Tan2013,He2013,Ge2015,Miyata2015,Shiogai2016} and heavily-doped LaFeAsO~\cite{Hiraishi2014}.

\begin{acknowledgments}
We thank K. Kimura, A. Shitade, and T. Yamashita for fruitful discussions. 
This work was supported by KAKENHI (Grants No. JP18H05227, No. JP18H01178, and No. JP20H05159) from the Japan Society for the Promotion of Science (JSPS). This work was supported by SPIRITS 2020 of Kyoto University.
\end{acknowledgments}

\appendix
\section{Geometric contribution to thermodynamic EQMs enhanced by band degeneracy
\label{appendix:degeneracy}}
In the main text, we have shown that the band degeneracy enhances the geometric contribution to the thermodynamic EQMs. 
To show this explicitly, 
we discuss an alternative expression of Eq.~\eqref{eq:TEQMs}:
\begin{widetext}
\begin{eqnarray}
    Q_{ij} = \dfrac{1}{2}\int_{\rm BZ}\dfrac{d^dk}{(2\pi)^d}
    \sum_{mn}
    \left[
        V_{nm}^i(\bm k)V_{mn}^j(\bm k) + c.c.
    \right]
    \left[
    \dfrac{1-\delta_{nm}}
    {\{\epsilon_n(\bm k)-\epsilon_m(\bm k)\}^2}
    \mathcal{F}_{nm}(\bm k)
    +\dfrac{\delta_{nm}}{12}
    f^{\prime\prime}(\epsilon_m(\bm k))\right]
    ,\label{eq:TEQMs_alt}\\
    \mathcal{F}_{nm}(\bm k) = 
        \dfrac{f(\epsilon_n(\bm k))+f(\epsilon_m(\bm k))}
        {2}
        -\dfrac{1}{\epsilon_n(\bm k)-\epsilon_m(\bm k)}
        \int_{\epsilon_m(\bm k)}^{\epsilon_n(\bm k)}
        d\epsilon f(\epsilon), 
    \label{eq:TEQMs_expand}
\end{eqnarray}
\end{widetext}
where $V_{nm}^i(\bm k) =\bra{u_n(\bm k)} \partial_{k_i}H(\bm k) \ket{u_m(\bm k)}$.
Since $\mathcal{F}_{nm}(\bm k) = O(1)$, the contribution from the non-degenerate bands is suppressed by the factor $\{\epsilon_n(\bm k)-\epsilon_m(\bm k)\}^{-2}$. 
For nearly degenerate bands, $|\epsilon_n(\bm{k})-\epsilon_m(\bm{k})| \ll T$, Eq.~\eqref{eq:TEQMs_expand} is approximated as
\begin{widetext}
\begin{align}
    \mathcal{F}_{nm}(\bm k) = 
        \dfrac{f(\epsilon_n(\bm k))+f(\epsilon_m(\bm k))}
        {2}- 1 
    +\dfrac{1}{\beta}\dfrac{\ln(1+e^{\beta \epsilon_n(\bm k)})-\ln(1+e^{\beta \epsilon_m(\bm k)})}
    {\epsilon_n(\bm k)-\epsilon_m(\bm k)}
    &\approx&\dfrac{1}{12}f^{\prime\prime}(\epsilon_m(\bm k))\left\{\epsilon_n(\bm k)-\epsilon_m(\bm k)\right\}^2.
\end{align}
\end{widetext}
Because the expression contains the factor $f^{\prime\prime}(\epsilon_m(\bm k))$, we notice that a large contribution is given by the momentum near the Fermi surface. From these discussions, we understand that the geometric contributions to the EQMs are enhanced by the band degeneracy near the Fermi level.
When all the bands are nearly degenerate, we have 
\begin{align}
    Q_{ij} = &\dfrac{1}{2}\int_{\rm BZ}\dfrac{d^dk}{(2\pi)^d}
    \sum_{mn}
    \left[
        V_{nm}^i(\bm k)V_{mn}^j(\bm k) + c.c.
    \right]
    \notag \\ &\times
    \left[
    (1-\delta_{nm})
    \dfrac{f^{\prime\prime}(\epsilon_m(\bm k))}{12}
    +\dfrac{\delta_{nm}}{12}
    f^{\prime\prime}(\epsilon_m(\bm k))\right].
\end{align}
In this case, the Fermi-sea term naturally has a similar form to the Fermi-surface term. In the realistic situation, the magnitude of the Fermi-sea term depends on the details of the electronic structure.

\section{Chemical potential dependence of the EQM in the 10-orbital model for LaFeAsO}\label{appendix:LaFeAsO_mu_dep}

\begin{figure}[htbp]
    \centering
    \includegraphics[width=0.8\linewidth]
    {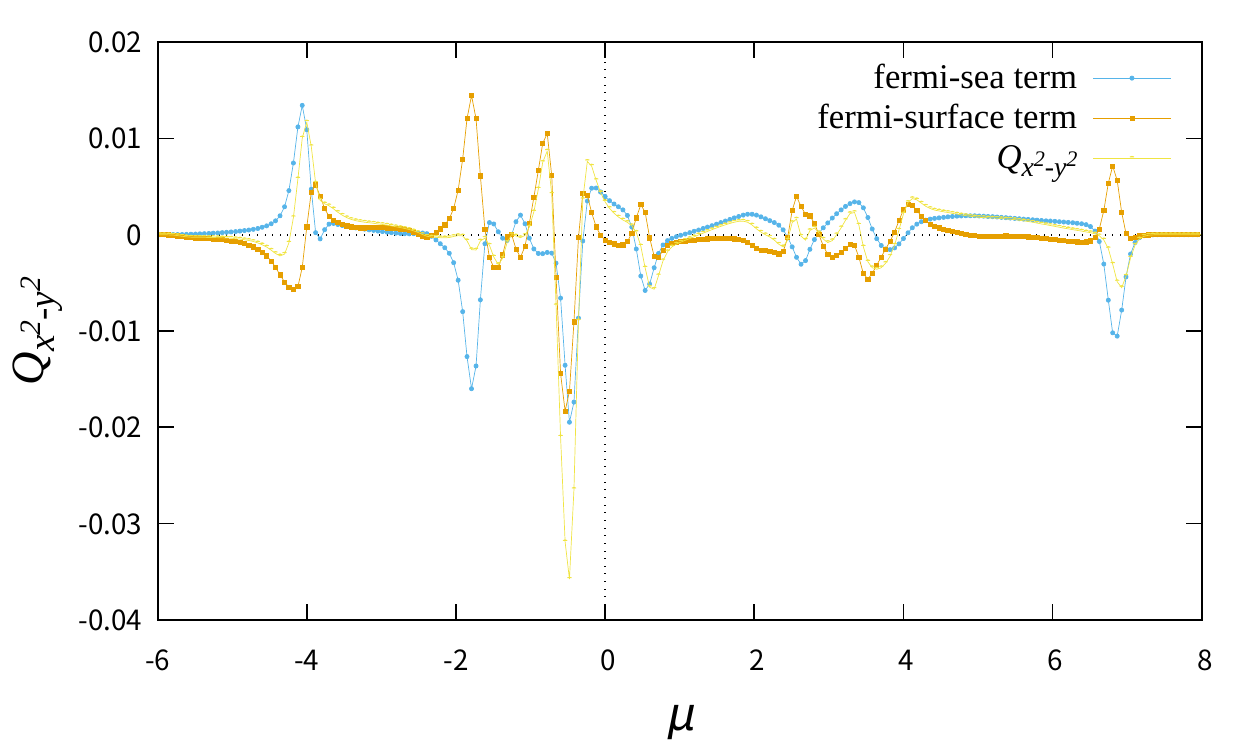}
    \caption{Chemical potential dependence of the thermodynamic EQM in the model for the tetragonal LaFeAsO. We set $T = 0.06$.
}
    \label{fig:LaFeAsO_tetra_mudep}
\end{figure}
Here we show the thermodynamic EQMs for various chemical potentials in the model of tetragonal LaFeAsO. Although in the main text the chemical potential is determined so that the particle number is 6.0, we change the chemical potential with keeping the other parameters. 
Figure~\ref{fig:LaFeAsO_tetra_mudep} is the result at $T = 0.06$.
It is shown that the geometric contribution is dominant only in a small part of the parameter range.
Thus, we consider that the dominant geometric contribution is a unique property of the iron-based superconductors for particle numbers around 6.0.
Actually, the nematic order of LaFeAsO$_{1-x}$F$_x$ has been observed in the region, $6.0 \leq n \leq 6.05$. 
We confirmed that the geometric term is dominant in this region.

\section{Tight-binding model reproducing Fermi surfaces of FeSe\label{appendix:FeSe_FS}}
To reproduce the Fermi surfaces of FeSe observed in experiments, we slightly modify the hopping parameters given by the first-principles calculation~\cite{ohnari_sign,Yamakawa2016,Ishizuka2018}. 
For this purpose, the energies of the $d_{xy}$-orbital band and the $d_{xz/yz}$-orbital band are shifted by ($-0.28,0,0.20$) and ($-0.27,0,0.13$) at  ($\Gamma$,$X$,$M$) points in the folded Brillouin zone, respectively. 
For this energy shift, the hopping parameters are changed so as to satisfy 
\begin{align}
    \delta E_{l} (\Gamma) &= \delta t_{ll}^{\rm on-site} +
    4\delta t_{ll}^{\rm nn} + 4\delta t_{ll}^{\rm nnn},\\
    \delta E_l (X) &= \delta t_{ll}^{\rm on-site},\\
    \delta E_l (M) &= \delta t_{ll}^{\rm on-site} -
    4\delta t_{ll}^{\rm nnn}, 
\end{align}
where we represent the energy shifts of the $l$-orbital band at $\Gamma$, $X$ and $M$ points
as $\delta E_l(\Gamma), \delta E_l(X)$ and $\delta E_l(M)$, respectively. The modification in the intra-orbital hopping integral is represented by 
$\delta t_{ll}$, and "on-sine", "nn", and "nnn" denote
the on-sine, first nearest neighbour, and second nearest neighbour hoppings, respectively.
In the 10-orbital model with two sublattices in the unit cell, $\delta t_{ll}^{\rm nn}$ ($\delta t_{ll}^{\rm nnn}$) is the inter-sublattice (intra-sublattice) hopping. 
We also tune the chemical potential to keep the filling $n=6$. 
Using these parameters, we obtain the Fermi surfaces in Fig.~\ref{fig:FeSe_FS_all}(a).

\section{Sign-reversing orbital polarization in FeSe\label{appendix:FeSe_orbital}}

\begin{figure}[htbp]
    \centering
    \includegraphics[width=0.9\linewidth]{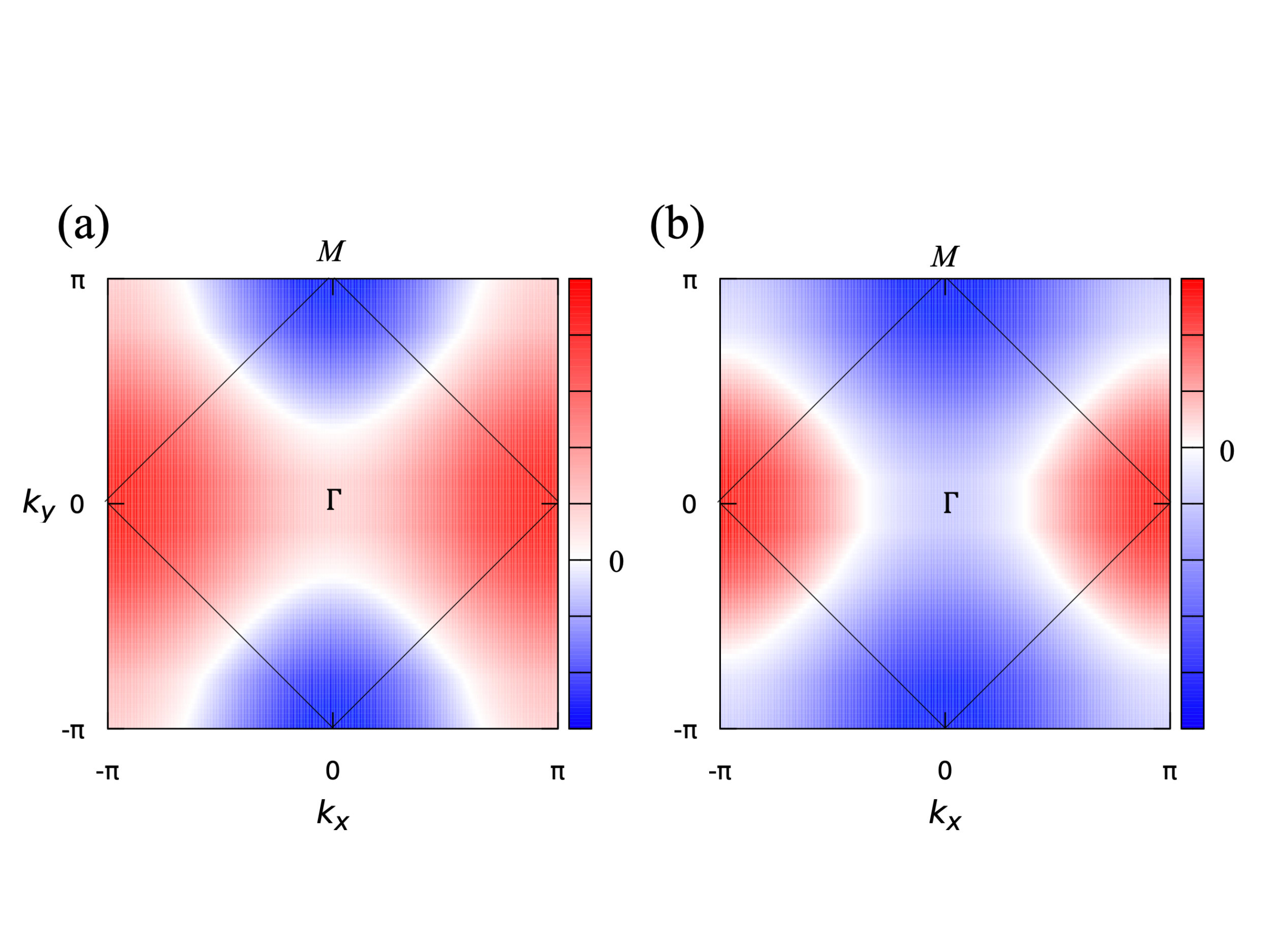}
    \caption{Momentum dependence of the energy shift due to the molecular field for FeSe, Eqs.~\eqref{eq:FeSe_Hamiltonian}-\eqref{eq:FeSe_bond}, in the unfolded BZ.
    (a) and (b) show $\delta E^{\,\rm nem}_{d_{xz}}(\bm k)$
    and $\delta E^{\,\rm nem}_{d_{yz}}(\bm k)$, respectivley.
    Red, blue, and white represent
    positive, negative, and zero value,
    respectively.
    The black lines show the BZ of the two-sublattice model.
    $\Gamma$ and $M$ are the points of the two-sublattice model.
    }
    \label{fig:dE_all}
\end{figure}

ARPES measurements clarified sign-reversing orbital polarization in FeSe~\cite{nakayama_sign,Shimojima_sign,Suzuki_sign,watson_sign,zhang_sign,Zhang_ARPES,Maletz_ARPES}, different from LaFeAsO. 
Thus, we introduce the molecular field, Eqs.~\eqref{eq:FeSe_Hamiltonian}-\eqref{eq:FeSe_bond}, yielding the orbital polarization with opposite sign between the $\Gamma$ and $M$ points. 
To understand the sign reversal, we here consider the unfolded BZ, for simplicity.
By Eqs.~\eqref{eq:FeSe_Hamiltonian}-\eqref{eq:FeSe_bond}, the molecular field gives the momentum-dependent energy shift
of the $d_{xz}$ and $d_{yz}$ orbitals as
\begin{align}
    \delta E^{\,\rm nem}_{d_{xz}}(\bm k) &=2\Delta(T) \left(\cos k_y  - \cos k_x  + \frac{1}{2}\right),\\
    \delta E^{\,\rm nem}_{d_{yz}}(\bm k) &=2\Delta(T) \left(\cos k_y - \cos k_x - \frac{1}{2}\right).
\end{align}
The momentum dependence in the energy shift is shown in Fig.~\ref{fig:dE_all}, which resembles a theoretical result for the sign-reversing orbital polarization~\cite{ohnari_sign}. 
In the folded Brillouin zone, ${\bm k}=(\pi,0)$ and $(0,\pi)$ are equivalent ($M$ point). Therefore, the energy splitting between the orbitals is $\delta E^{\,\rm nem}_{d_{xz}}(0,\pi)-\delta E^{\,\rm nem}_{d_{yz}}(\pi,0)$ at the $M$ point, while it is $\delta E^{\,\rm nem}_{d_{xz}}(0,0)-\delta E^{\,\rm nem}_{d_{yz}}(0,0)$ at the $\Gamma$ point. As shown in Fig.~\ref{fig:dE_all}, the sign is opposite between the $\Gamma$ and $M$ points, consistent with the sing-reversing orbital polarization in FeSe~\cite{nakayama_sign,Shimojima_sign,Suzuki_sign,watson_sign,zhang_sign,Zhang_ARPES,Maletz_ARPES}.

\section{EQM in three-orbital $d$-$p$ model for cuprate superconductors\label{appendix:dp-model}}

Here, we show the EQM in the 3-orbital $d$-$p$ model which has been studied for cuprate superconductors~\cite{Luo-Bickers1993,Koikegami1997,Takimoto1997}.
The model takes into account the $d_{x^2-y^2}$-orbital of coppers and the $p_x$ and $p_y$ orbitals of oxygens.
For comparison with the $17$-orbital model studied in the main text, we assume the half filing. The hopping parameters and the molecular field are the same as those in the 17-orbital model.
In the cuprates, there is no band degeneracy near the Fermi surface, and the low-energy electron states are appropriately described by the 3-orbital $d$-$p$ model. Thus, when the EQMs are mainly given by the Fermi-surface term, we expect qualitatively the same results as the $17$-orbital model.

\begin{figure}[htbp]
    \centering
    \includegraphics[width=0.8\linewidth]
    {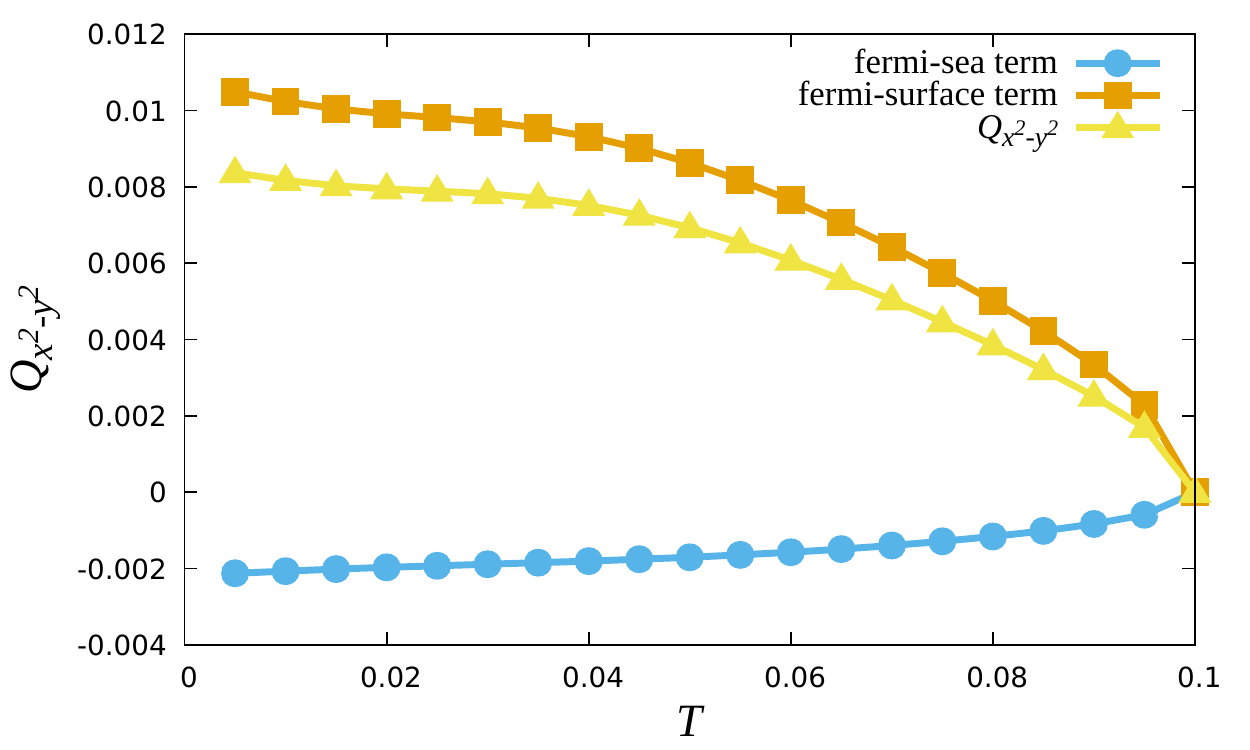}
    \caption{Thermodynamic EQMs in the 3-orbital $d$-$p$ model for cuprate superconductors with bond order.
}
    \label{fig:La2CuO4_bond_EQMs}
\end{figure}

\begin{figure}[htbp]
    \centering
    \includegraphics[width=0.8\linewidth]
    {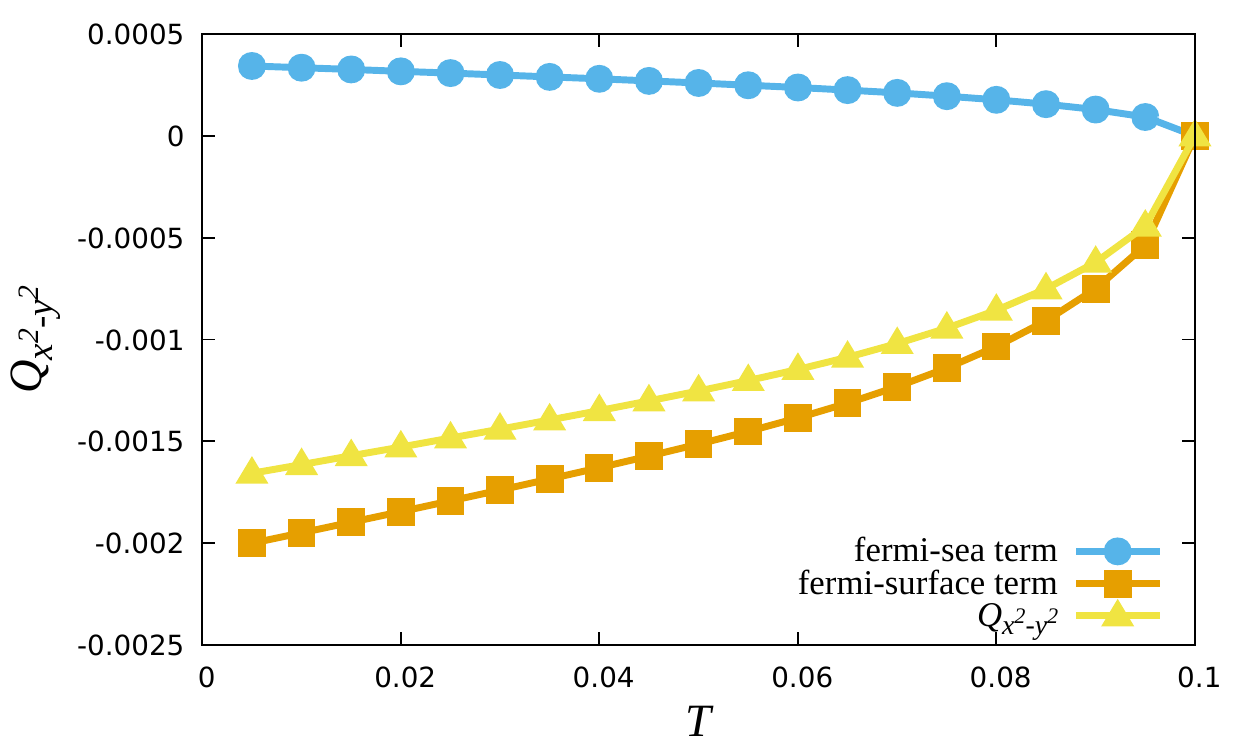}
    \caption{Thermodynamic EQMs in the 3-orbital $d$-$p$ model for cuprate superconductors with orbital order.
}
    \label{fig:La2CuO4_orbital_EQMs}
\end{figure}

Indeed, the thermodynamic EQMs show the similar behaviors to those in the $17$-orbital model, as shown in Figs.~\ref{fig:La2CuO4_bond_EQMs} and \ref{fig:La2CuO4_orbital_EQMs}.
In both cases of the $d_{x^2-y^2}$-wave bond order and the $p$-orbital order, the Fermi-surface term is dominant. 
The magnitude of the EQMs is larger in the bond-ordered state than the orbital-ordered state, like in the $17$-orbital model. 
On the other hand, we see differences in the Fermi-sea term between
the $3$-orbital and $17$-orbital $d$-$p$ models: even the sign is opposite in the orbital-ordered state. This implies the importance of the realistic multi-orbital model for the evaluation of Fermi-sea terms.

\bibliography{main}

\end{document}